\documentclass[prb,nobibnotes,aps,unsortedaddress,showkeys]{revtex4} 

\usepackage{times,helvet}

\usepackage{graphicx}
\usepackage{subfigure}
\usepackage{afterpage} 
\usepackage{color}

\usepackage{amsmath}
\usepackage{amssymb}
\usepackage{dsfont}
\usepackage{mathrsfs}

\usepackage{fancyhdr}

\makeatletter
\renewcommand\@dotsep{2}
\makeatother

\begin{document}

\title{A variational approach to the stochastic aspects of cellular signal
  transduction }

\author{Yueheng Lan}
\affiliation{Department of Chemistry, University of North Carolina at Chapel Hill, NC 27599-3290}

\author{Peter G. Wolynes}
\affiliation{Department of Chemistry \& Biochemistry, University of California at San Diego, 9500 Gilman Dr., La Jolla, CA 92093-0371}

\author{Garegin A. Papoian}
\email[Electronic mail: ]{gpapoian@unc.edu}
\affiliation{Department of Chemistry, University of North Carolina at Chapel Hill, NC 27599-3290}

\date{\today}

\begin{abstract}

  Cellular signaling networks have evolved to cope with intrinsic
  fluctuations, coming from the small numbers of constituents, and the
  environmental noise. Stochastic chemical kinetics equations govern the way
  biochemical networks process noisy signals. The essential difficulty
  associated with the master equation approach to solving the stochastic
  chemical kinetics problem is the enormous number of ordinary differential
  equations involved. In this work, we show how to achieve tremendous
  reduction in the dimensionality of specific reaction cascade dynamics by
  solving variationally an equivalent quantum field theoretic formulation of
  stochastic chemical kinetics. The present formulation avoids cumbersome
  commutator computations in the derivation of evolution equations, making
  more transparent the physical significance of the variational method.  We
  propose novel time-dependent basis functions which work well over a wide
  range of rate parameters. We apply the new basis functions to describe
  stochastic signaling in several enzymatic cascades and compare the results
  so obtained with those from alternative solution techniques. The variational
  ansatz gives probability distributions that agree well with the exact ones,
  even when fluctuations are large and discreteness and nonlinearity are
  important. A numerical implementation of our technique is many orders of
  magnitude more efficient computationally compared with the traditional Monte
  Carlo simulation algorithms or the Langevin simulations.
\end{abstract}


\preprint{Submitted to J. Chem. Phys.}

\maketitle 

\section{Introduction}

The life of the cell is regulated by intricate chains of chemical
reactions~\cite{st02gomp}. The whole cell may be viewed as a computing device
where information is received, relayed and processed~\cite{den95p}.  Signal
transduction cascades based on protein interactions regulate cell movement,
metabolism and division~\cite{st02gomp,mol02alb}. Since cells are mesoscopic
objects, understanding the role of the intrinsic fluctuations of the
biochemical reactions as well as environmental fluctuations is a fundamental
part of understanding signaling
dynamics~\cite{paul00st,tan04mod,sh05noise,bark99cir,kurt95st,
  el02st,sw02in,oz02reg,bla03noi,wein05st,muk04st,rao02n}. In this regard, the
well-organized behavior of cells, which emerges as a result of biochemical
reaction dynamics involving hundreds of cross-linked signaling pathways, is
remarkable~\cite{han01ex,nick04ph,h03ecoli,coh05fl,
  keen01df,lem05st,kul04pat}. The problem of how signals can be precisely
detected, smoothly transduced and reliably processed under noisy conditions is
a research topic of great current interest, that, in turn, should lead to
deeper understanding of the origins of the cell's functional
responses~\cite{math02hei,opt04cha}. Furthermore, these studies can help to
unravel the design principles for various signaling pathways, leading,
eventually, to better ways to control and efficiently interfere with cellular
activity, as would be needed to correct the behavior of diseased
cells~\cite{keen05len,h03ecoli}.

The role of noise in gene regulatory networks has been identified as a key
issue and has been intensively studied in recent
years~\cite{th01in,kie01eff,th01st,sw02in,oz02reg,bla03noi,sasai03stch,walcz04stgn,jas04fl}.
Linearization of the noise may be acceptable if the dynamics near steady
states is being studied~\cite{th01in,jas04fl,sw02in}.  When protein numbers
are large and, thus, the continuous approximation is valid, time-dependent
distributions have been determined using the Langevin or Fokker-Planck
equations ~\cite{muk02att,sw04eff,sh05noise}.  To account for the discreteness
in the linearized equations, the generating function approach has also been
used~\cite{th01in,sw02in}. A variational treatment of steady state stability
and switching in nonlinear, discrete gene regulatory processes has been
reported\cite{sasai03stch,walcz04stgn}.

In cytosolic signal transduction processes, in contrast to gene transcription
which involves a unique DNA molecule, all the reacting species are
present in multiple copies and participate in unary, binary or perhaps even
higher order reactions. Noise could be multiplicative~\cite{van92st,fox86fun}
and the linear description easily breaks down. Moreover, cellular reactions
usually take place heterogeneously in space The localization and
compartmentalization of protein organelles require diffusive or active
transportation of reacting molecules from one region to another. Spatial
coordination combined with temporal coordination generates coherent, yet
complex spatiotemporal
patterns~\cite{hol05st,coh05fl,ku01sel,keen01df,lem05st,kul04pat,
  h03ecoli,sh02mod}.

The extracellular ligands often trigger cascades of chemical reactions which
propagate inside a cell and induce responses from various environmental
cues. The cell body is a highly heterogeneous entity and never settles to a
steady state. To understand cell dynamical processes, an explicitly
time-dependent description is required. Within a volume with linear dimensions
of the Kuramoto length~\cite{van92st,kur74eff}, diffusion mixes the reagents
in a nearly uniform manner. If the reactions are considered in the Kuramoto
volume, it is reasonable to neglect the spatial heterogeneity. For many signal
transduction networks, however, it is likely that only a few proteins are
present in the Kuramoto volume (determined by specific reaction and diffusion
rates), and, therefore, the continuous description of protein numbers breaks
down. To characterize stochastic signaling reactions in this volume, a
time-dependent description of a noisy, discrete, nonlinear system is
required~\cite{lan1cas}. In many situations, such as {\em Drosophila}
oogenesis, the exact shape of the probability distribution profile is very important
and determines different developmental paths~\cite{sh02mod,yak05sys}. In the
following, we discuss efficient techniques to compute the time-dependent
protein number probability distributions in biochemical reaction networks when
the number of protein copies is small.

The Gillespie algorithm provides an effective Monte Carlo technique for
simulating stochastic chemical
reactions~\cite{gill77ext,gil01app,jer05sim}. Each simulation gives a reaction
trajectory which is close to the deterministic trajectory in the large
particle number limit.  To get well-converged statistics, many trajectories
may be needed, often on the order of $10^5$.  If there is a separation of time
scales of the constituent chemical reactions, Gillespie simulations also
become exceedingly slow since the reaction events are dominated by the fastest
reactions while the observables typically involve the slowest
reactions. Although considerable progress has been made in accelerating such
simulations~\cite{e05nest,muk02att,sw04eff,sh05noise}, computational
inefficiency continues to be an impediment, especially for the spatially
inhomogeneous generalization of the Gillespie algorithm. Furthermore, it is
hard to extract the analytical structure of the solution from the numerical
results, which can be important for achieving a deeper physical understanding
of the system behavior when the rate parameters are widely varied.

Mathematically, a stochastic process may be completely characterized by a
master equation - a group of ordinary differential equations (ODEs) describing
the evolution of probabilities~\cite{van92st,gar02han}. The main difficulty in
solving a chemical master equation is the enormous number of ODEs involved
even for a small reaction cascade. A number of analytical techniques have been
developed for solving approximately the master
equation~\cite{th01st,th01in,sw02in}. In this work, we show how to achieve
enormous reduction in the dimensionality of specific reaction cascade dynamics
by solving variationally the quantum field theory (QFT) equations of
stochastic chemical
kinetics~\cite{doi76st,ons53fl,mat98qft,zin02qft,doi76sq}. Our present
approach is based on mapping the master equation ODEs into a single partial
differential equation (PDE) and applying a variational technique which reduces
the PDE into a small number of ODEs. The variational QFT approach has been
employed to study steady state stability and switching in gene regulatory
networks\cite{sasai03stch,walcz04stgn}.  In this work we propose novel
time-dependent basis functions appropriate for describing protein signaling
cascades which work in a wide range of rate parameters. Our method gives
probability distributions that agree well with the exact ones, including when
the fluctuations are large and discreteness and nonlinearity play important
roles.

The paper is organized as follows. In section \ref{sect:qft}, the QFT
formulation of the stochastic processes describing chemical
reactions~\cite{zin02qft,sasai03stch,walcz04stgn,doi76sq,doi76st,zel77fl} and Eyink's
variational solution technique for solving such field
theories~\cite{eyink96act,yink97r} are presented. We show that the QFT
formulation is equivalent to a generating function approach and also discuss
the physical significance of the variational principle in this context. In
section \ref{sect:num}, we apply the new trial functions and the variational
technique to a number of 2-step, 3-step, and 4-step enzymatic reaction
cascades and compare our results with those found with more traditional
methods. Finally, we discuss the more general principles of basis function
construction and the limitations of variational approaches.

\section{Quantum field theory formulation, variational principle and generating functions}
\label{sect:qft}

In this section, we first discuss briefly the master equation and demonstrate
its application to a 2-step signal amplification cascade. Next, the master
equation is recast into a QFT form in which the probability evolution is
governed by a ``wave equation''. Then, we show that the field theoretic
formulation is equivalent to a generating function approach. To solve these
equations, Eyink's variational technique and its physical significance are
examined. We further explore the practical implementation issues in section
\ref{sect:sum}.

\subsection{The master equation and its solution}

Unlike a stochastic simulation which produces an individual random trajectory
and generates statistics only after a large number of samplings, master
equations directly describe the evolution of probability distributions in the
state space of a system based on specific inter-state transition rates. For a
discrete system, the master equation consists of a set of ODEs (see the
following examples), while for a continuous system it becomes an
integro-differential equation such as the Boltzmann equation. Although the
master equation is a complete description of a Markovian system, its solution
is usually difficult and requires special techniques.  This paper presents one
variational technique that could be used.

\begin{figure}[h]
\includegraphics[width=14cm]{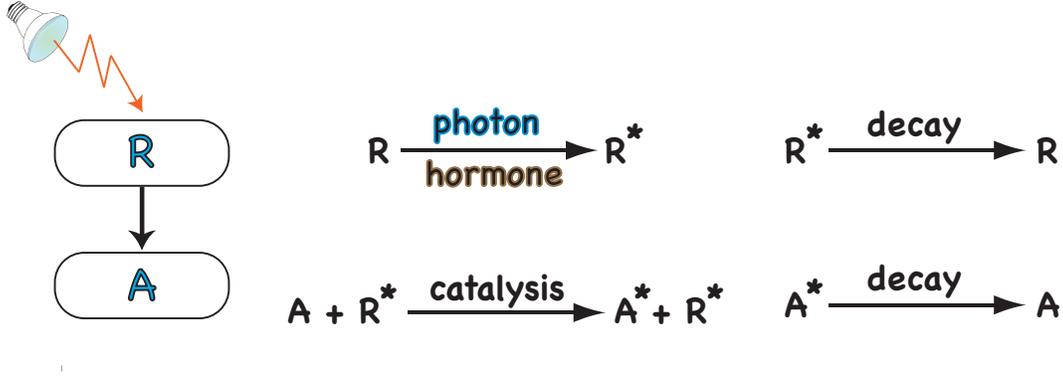}
\caption{An inactive receptor $R$, when activated by a signal, activates
  downstream protein $A$.}
\noindent 
\label{f:react}
\end{figure}

As an example, let's consider the following set of equations that represents
the simplest enzymatic signal amplification process (Fig. \ref{f:react}). In
this simple reaction scheme, without feedback loops, $R$ represents an
inactive receptor, which becomes activated into $R^*$ upon binding of an
external ligand (stimulus). When the receptor is activated, it acts as an
enzyme, catalyzing the phosphorylation of the next kinase downstream ($A+R \to
A^*+R^*$) with a rate $\mu$. $A^*$ spontaneously decays to $A$ with a rate
$\lambda$ and $R^*$ to $R$ with a rate $k$.  In the absence of $R^*$, $A \to
A^* $ may occur naturally, however, with a much lower rate, so that it can be
ignored when we introduce the catalyst $R^*$.

Although the $R^*$ reaction is unary and independent of the $A$ reaction, the
latter one is binary, making the system nonlinear, thus, different from those
considered in a number of prior works on the gene regulatory
networks\cite{th01in,kie01eff,sw02in,oz02reg}.  To write the master equation,
we denote by $P(m,n)$ the probability of having $m$ $R^*$'s and $n$ $A$'s,
then
\begin{eqnarray}
\frac{dP}{dt}(m,n) &=& \mu[-mnP(m,n)+m(n+1)P(m,n+1)] \nonumber \\ 
&+& \lambda[-(N-n)p(m,n) + (N-n+1)P(m,n-1)]  \nonumber \\
&+& g[-P(m,n)+P(m-1,n)]+k[-mP(m,n)+(m+1)P(m+1,n)]
\,, \label{eq:m1}
\end{eqnarray}    
where $N$ is the total number of $A$ and $A^*$. In Eq.~(\ref{eq:m1}), the
first two terms describe the $A-A^*$ reaction and the rest the $R-R^*$
reaction.  This simple 2-step cascade is commonly found embedded in the onset
of a reaction pathway of many important signaling
cascades\cite{pugh92r,sch02com}. If a large number of inactive receptors, $R$,
are present the rate of conversion depends on the arrival times of the
external cue and the reaction becomes Poissonian.  We assume that this is the
case in all the following calculations.  If the $R \to R^*$ reaction is the
usual birth-death problem, our formalism still applies with only minor
changes. The master equation~(\ref{eq:m1}) actually contains infinitely many
coupled ODEs.

\subsection{The QFT formulation}

The differential-difference equations, such as Eq. (\ref{eq:m1}), are well
represented in the QFT formulation by introducing creation and annihilation
operators $a,a^\dagger$ and states
$|n\rangle$~\cite{doi76st,ons53fl,mat98qft,zin02qft,doi76sq,sasai03stch,walcz04stgn}. In
analogy to quantum mechanics, the operators satisfy the commutation relation
that
\[
[a,a^\dagger]=1
\,.
\]
As usual, the ``vacuum state'' $|0\rangle$ and its conjugate $\langle 0|$ are
defined to satisfy
\[
\langle 0|a^\dagger=a|0\rangle=0\,, \quad \langle0|0\rangle=1
\,.
\]
Other states are built up from the vacuum state, such as the $n$-particle
state $|n\rangle$
\[
|n\rangle=a^{\dagger n}|0\rangle
\,.
\] 
It is easy to check with the help of the commutation relation
\[
a|n\rangle=n|n-1\rangle\,,\quad a^\dagger |n\rangle=|n+1\rangle \,, 
\quad a^\dagger a |n\rangle=n|n\rangle
\,.
\]
Hence, $a^\dagger a$ is the ``particle number operator''.  Notice that the states
$|n\rangle$ are not normalized in the usual sense since
\[
\langle n|n\rangle=\langle 0|a^n|n\rangle=n!
\,,
\] 
but they are orthogonal,
\[
\langle m|n\rangle=\langle0|a^m|n\rangle=0\,, \mbox{ for } m \neq n
\,.
\]
The state that corresponds to a probability distribution $P(n)$ is
\[
|\Psi\rangle=\sum_n P(n)|n\rangle
\,.
\]
The probabilities are thus encoded into the coefficients of different particle
number states superimposed into the ``wave function'' $|\Psi\rangle$.  In
order to compute physical observables, the {\em harvesting state}
$\langle\phi|=\langle0|e^a$ is introduced. It is easy to check that
\[
\langle\phi|n\rangle=1\,,\quad \langle\phi|\Psi\rangle=1\,,\quad \langle\phi|(a^\dagger a)^m|\Psi\rangle
=\langle n^m\rangle
\,.
\]
The first equation shows the particular normalization of an $n$-particle
state.  The second equation corresponds to the probability conservation
$\sum_n P(n)=1$.  The third equation may be used to calculate the $m$-th
moment of the particle number.  The evolution of probabilities is governed by
a wave equation for $\Psi$:
\begin{equation}
\frac{d|\Psi\rangle}{dt}=\Omega |\Psi\rangle
\,. \label{eq:mg}
\end{equation}
The original large sets of ODEs are now compactified into just one equation.
Applied to the 2-step cascade (Fig. \ref{f:react}), Eq. (\ref{eq:mg}) is
characterized by following operator $\Omega$,
\begin{equation}
\Omega=(1-a^\dagger)(\mu b^\dagger b a-\lambda N+\lambda a^\dagger a)
+g(b^\dagger -1)+k(b-b^\dagger b)
\,, \label{eq:m4om}
\end{equation}
where $b^\dagger\,,b$ are the creation and annihilation operators associated
with $R^*$ and $a^\dagger\,,a$ with $A$.  In this case,
\[
|\Psi\rangle=\sum_{m,n} P(m,n)|m,n\rangle
\,,
\]
where
\[
|m,n\rangle=a^{\dagger m}b^{\dagger n}|0\rangle
\,.
\] 
Eq.~(\ref{eq:m4om}) is readily verified by substituting into Eq.~(\ref{eq:mg})
and comparing the coefficients of each $(m,n)$-particle state.  In contrast to
ordinary quantum mechanics, the operator $\Omega$ is non-Hermitian, so the
inner products between the states are not conserved. This was the reason to
introduce earlier the harvesting state. Nevertheless, many QFT techniques may
be profitably applied, albeit with some
modifications~\cite{ons53fl,wang94sur,zin02qft,sasai03stch,walcz04stgn,shnerb01aut,Krish03phas}.
We will not discuss those and instead will translate the above field theoretic
formulation to the familiar differential equation language.

\subsection{Differential operators and the generating function}

In the field theoretic form, the computations are carried out by commutator
manipulations that sometimes are awkward. Fortunately, it turns out that we
may convert the operator equation (\ref{eq:mg}) into a partial differential
equation (PDE). To accomplish that, we explore the analogy between
$a\,,a^\dagger$ and $d/dx \,,x$. Not only do they have the same commutator
\[
[a,a^\dagger]=1 \Longleftrightarrow \quad [\frac{d}{dx},x]=1
\,,
\]
but, a more comprehensive correspondence is found:
\begin{eqnarray*}
aa^\dagger |0>=|0\rangle  \Longleftrightarrow \quad \frac{d}{dx} x=1 \,; \\
a|0\rangle=0 \Longleftrightarrow\quad \frac{d}{dx}1=0 \,;\\
aa^{\dagger n}|0\rangle=n a^{\dagger n-1}|0\rangle \Longleftrightarrow \frac{d}{dx} x^n=nx^{n-1};\\
a^m a^{\dagger n}|0\rangle=\frac{n!}{m!}a^{\dagger n-m}|0> \Longleftrightarrow
(\frac{d}{dx})^m x^n=\frac{n!}{m!}x^{n-m}\,,\mbox{ for } n \ge m
\,.
\end{eqnarray*}
From these, we can also deduce for any smooth function $f$ that
\begin{eqnarray*}
af(a^\dagger)|0\rangle \Longleftrightarrow \frac{d}{dx}f(x) \,;\\
a^mf(a^\dagger)|0\rangle \Longleftrightarrow (\frac{d}{dx})^mf(x)
\,.
\end{eqnarray*}
The analogy $|\Psi\rangle=\sum_n P(n) a^{\dagger n}|0 \rangle
\Longleftrightarrow \Psi(x)=\sum_n P(n) x^n$ converts a wavefunction to a
generating function.  The inner product with the harvesting state corresponds
to evaluation at $x=1$.  It is easy to check the following relations,
\begin{eqnarray*}
<0|e^a|\Psi> &=&\Psi(1)\,;\\
<0|e^af(a^\dagger)|\Psi>&=& f(1)\Psi(1)\,,\\
\mbox{as }  e^{d/dx}f(x)&=& f(x+1)
\,.
\end{eqnarray*}
The wave equation (\ref{eq:mg}) becomes then a PDE for the generating function
after all the necessary conversions are done. For the 2-step cascade, this
expression is
\begin{eqnarray}
\frac{\partial \Psi}{\partial t} &=& (1-x)(\mu y\frac{\partial^2}{\partial
  x \partial y}-\lambda N+\lambda x \frac{\partial}{\partial x})\Psi
  +g(y-1)\Psi-k(y-1)\frac{\partial \Psi}{\partial y} \label{eq:m4dx}
\,,
\end{eqnarray}
where the first term describes the $A-A^*$ reaction and the rest the $R-R^*$
reaction. Generating functions were previously used to treat unary reactions
in the gene regulatory network~\cite{ott00fl,sw02in}. The second order
derivative term $\partial^2 \Psi/\partial x\partial y$ in Eq.(\ref{eq:m4dx})
is characteristic of a binary reaction, indicative of nonlinear kinetics. This
higher derivative changes the order of the PDE and adds significant difficulty
to solving Eq.(\ref{eq:m4dx}).  In the generating function formalism, the
equation $\Psi(1)=1$ encodes the conservation of probability and the first two
moments are given by
\begin{eqnarray}
\langle n \rangle &=& \frac{\partial \Psi}{\partial x}|_{x=1} \,,\nonumber \\
\langle n^2 \rangle &=& (\frac{\partial^2 \Psi}{\partial x^2}+
\frac{\partial \Psi}{\partial x})|_{x=1}
\,, \label{eq:genm}
\end{eqnarray}
where $|_{x=1}$ means evaluation at $x=1$. Therefore, the moments are obtained
when the generating function is expanded at $x=1$, while the probability
distribution is obtained from the Taylor coefficients when the generating
function $\Psi$ is expanded at $x=0$.

\subsection{The variational solution}

In the QFT formulation of the stochastic processes, a variational principle
may be derived which is equivalent to the evolution equation
(\ref{eq:mg}). This principle indicates that the physical solution of
Eq.(\ref{eq:mg}) is given by the stationary points of the following
functional~\cite{sasai03stch,walcz04stgn,eyink96act}
\begin{equation}
H[\Phi_L\,,\Phi_R]=\int_0^{\infty}dt\langle\Phi_L|\partial_t-\Omega|\Phi_R\rangle
\,,\label{eq:var1}
\end{equation}
where $\Phi_L$ and $\Phi_R$ are arbitrary quantum states under quite general
constraints consistent with the positivity of probabilities and the fixed
boundary conditions. In practice, we take a finite-dimensional subset of the
infinite-dimensional function space and apply the variational principle in
this subspace to get closed equations that may be subsequently solved by
simple numerical calculation. If the essential qualitative properties of the
system are known, good approximations of the original problem can be achieved
through an informed choice of time-dependent basis functions that define the
relevant subset in the function space.

Because $\Omega$ is not Hermitian, the right and left eigenvectors are
different.  To characterize the system, we, therefore, need two sets of
vectors $\Phi_L$ and $\Phi_R$.  The stationary variation condition for
$\Phi_L$ restores the original equation (\ref{eq:mg}) and that for $\Phi_R$
defines an equation satisfied by $\Phi_L$.  If we view the operator
$\partial_t - \Omega$ as a large matrix parameterized by $t$, the $\Phi_L$ and
$\Phi_R$ generated by the stationary variation condition correspond to its
singular vectors~\cite{nr,gol96mat} and the extremum values of
Eq.~(\ref{eq:var1}) are the singular values. Physically, from the
Schr{\"o}dinger picture point of view, $\Phi_R$ is the evolving quantum state
and $\Phi_L$ represents the measurable quantities in which we are
interested. Eq.~(\ref{eq:var1}) serves to find the most significant state and
physical observables. Eyink originally applied this variational principle to
Fokker-Planck equations~\cite{yink97r}. Subsequently, Sasai and Wolynes used
this variational approach in the field theoretic form and obtained moments in
a toggle-switch gene regulatory problem~\cite{sasai03stch}. In this paper, we
show how the variational principle may be applied, instead, to the generating
functions. We introduce novel basis functions to obtain the time-dependent
probability distributions in signal transduction cascades. Another novelty of
the present formulation is our avoidance of cumbersome commutator computations
in the derivation of the evolution equations, making more transparent the
physical significance of the variational method.

There are many ways to choose the time-dependent basis functions. We follow
the approach of Sasai and Wolynes~\cite{sasai03stch}:
\begin{eqnarray}
|\Phi_R \rangle &=& \Phi_R(a^{\dagger},\{f_i(t)\}_{i=1}^n)|0\rangle \label{eq:varr} \\
\langle \Phi_L| &=& \langle 0| \exp(a)(1+\sum_{i=1}^m c_i(t) a) \label{eq:varbl}
\,,
\end{eqnarray}
where $n$ is the number of unknown functions in $|\Phi_R\rangle$ and $m$ is
the number of parameters in Eq.~(\ref{eq:varbl}). The exponential factor in
$\langle \Phi_L|$ acting on $\langle 0 |$ gives the harvesting state. If we
substitute this ``ansatz'' into Eq.(\ref{eq:var1}) and carry out the variations
with respect to $c_i$, a finite set of ODEs for the evolution of
$\{f_i(t)\}$ are obtained, which then determines the evolution of the probability
distribution.

As mentioned, the variational method can also be recast into the generating
function language using the conversion scheme discussed previously.  Now
$\Phi_L$ becomes a differential operator and $\Phi_R$ a guess function of
variable $x$.  For example, Eq.~(\ref{eq:varr},\ref{eq:varbl}) corresponds to
\begin{eqnarray}
\Phi_R &=& \Phi_R(x,\{f_i(t)\}) \,, \nonumber \\
\Phi_L &=& 1+\sum_{i=1}^m c_i(t) \frac{d^i}{dx^i}
\,. \label{eq:varb2}
\end{eqnarray}
The functional $H$ simply becomes
\begin{equation}
H=\int_0^{\infty}dt\Phi_L(\partial_t-\Omega)\Phi_R |_{x=1}
\,,\label{eq:varb3}
\end{equation}
In the new picture, we have much simpler mathematical operations, {\em e.g.},
the variational principle becomes simply a function extremization condition
\begin{equation}
\frac{\delta H}{\delta c_i(t)}|_{\{c_j(t)=0\}_{j\leq m},x=1}=0\,, \mbox{ for } i=1,2,\dots,m
\,.\label{eq:varg1}
\end{equation}
Or equivalently
\begin{equation}
\frac{d^i}{dx^i}(\partial_t-\Omega)\Phi_R|_{x=1}=0\,, , \mbox{ for } i=1,2,\dots,m
\,.\label{eq:varg2}
\end{equation}
The evolution of the generating function should always conserve the total
probability. As in Eq.~(\ref{eq:m4dx}), the total probability $\Psi(1,1)$ does
not change with time. The proper choice of $\Phi_R$ should also guarantee this
invariance, satisfying $(\partial_t-\Omega)\Phi_R|_{x=1}=0$. Now,
Eq.~(\ref{eq:varg2}) tells us that the higher derivatives of the expression
$(\partial_t-\Omega)\Phi_R$ evaluated at $x=1$ are also zero. Therefore, in
the limit of $m \to \infty$, Eq.~(\ref{eq:varg2}) leads to the PDE $\partial_t
\Psi_R=\Omega \Psi_R$. For finite $m$, this PDE is approximately satisfied in
the neighborhood of $x=1$.

\section{Numerical applications}
\label{sect:num}

In this part, we discuss the implementation of the PDE version of the
variational method and apply it to several simple, yet important enzymatic
cascades. Before proceeding to the individual examples, we emphasize our
motivation for selecting the time-dependent basis functions. We also briefly
discuss several alternative methods also used to solve the master equation.

\subsection{Computational details}
It is reasonable to require the following constraints on the right basis function $\Phi_R(x,y)$:\\
(1) the total probability should be equal to 1, {\em i.e.}, $\Phi_R(1,1)=1$;\\
(2) the probability should be positive, {\em i.e.}, the coefficients of the
Taylor expansion of $\Phi_R$ around $(x,y)=(0,0)$ should be nonnegative;\\
(3) the time rate of the unknown functions, ${\dot{f}_i(t)}$,
should be obtainable by solving
Eq.~(\ref{eq:varg2}) derived from the variational principle.\\
In the following,
we introduce two sets of basis functions. One set is simple but is of limited
applicability, while the other is in a more complex integral form and can be
applied very generally.

We use simple left basis functions, $\Phi_L(x,y)$. As suggested in
Eq.~(\ref{eq:varg1}), they are represented by differential operators and can
be easily extended to multi-variate cases.  For the 2-step cascade, we use
\begin{equation}
\Phi_{L,1}=1+c_1\partial_x+c_2\partial_y
 \label{eq:cas2dl1}
\end{equation}
with a simple right basis function (see Eq. (\ref{eq:ansatz1}) below) and
\begin{equation}
\Phi_{L,2}=\Phi_{L,1}+c_3\partial_{yy}
\label{eq:cas2dl2}
\end{equation}
with a more complicated right basis (see Eq.~(\ref{eq:ans2}) below). For the
3-step cascade discussed below, we use
\begin{equation}
\Phi_{L,3}=\Phi_{L,2}+c_4\partial_{z}+c_5\partial_{zz}
\,. \label{eq:cas3dl}
\end{equation}
Following a similar pattern, we can simply write the left basis function for
the 4-step cascade discussed below, as,
\begin{equation}
\Phi_{L,4}=\Phi_{L,3}+c_6\partial_{w}+c_7\partial_{ww}
\,. \label{eq:cas4dl}
\end{equation}
In all above equations, $\partial_x\,,\partial_{xx}$ denote the first and
second derivative with respect to $x$, and so on. Other choices of the left
basis function are, of course, possible. The current choice is simple and gave
the best results among different basis functions which we tried. ``Maple''
symbolic software was used to derive the time evolution equations and,
subsequently, ``Matlab'' numerical software was used to carry out the
evolution of equations of motion and for plotting the computation results.

To validate our calculations, we used the Gillespie
simulation~\cite{gill77ext,gil01app,jer05sim,tan04mod} results as the
reference (exact) solutions. $10^5$ stochastic trajectories were sampled to
derive the distributions and other statistical quantities. At the same time,
the variational method was also compared with two commonly used methods -
Langevin equation~\cite{van92st} and
$\Omega$-expansion~\cite{van92st,elf03fast,lin04hay,tao05st}. In the Langevin
equation simulation, we also used $10^5$ realizations. In addition, to prevent
the appearance of negative particle numbers, we applied the selection
procedure commonly used~\cite{fox94em}. It is awkward and time-consuming to
use the $\Omega$-expansion method to compute the distributions. We only
applied it to the simplest 2-step cascade.

\subsection{Application to a two-step amplification cascade}

In a previous paper~\cite{lan1cas}, approximate analytical solutions for
(\ref{eq:m4dx}) in certain parameter range were obtained using the method of
characteristics.  If the initial conditions correspond to zero $R^*$'s and N
$A$'s, then a generating function solution reads,
\begin{equation}
\Psi(x,y) = [1+\left(\frac{\lambda}{\lambda+\mu m(t)}
+\frac{\mu m(t)}{\lambda+\mu m(t)}e^{-(\lambda+\mu m(t))t}\right)(y-1)]^N 
\exp(m(t)(x-1))
\,, \label{eq:phi0th1} 
\end{equation}
where $m(t)=\frac{g}{k}(1-e^{-kt})$ is the average number of $R^*$ at time $t$.
We make use of this specific functional form and try the following ansatz,
\begin{equation}
\Phi_R = (1+f_2(t)(y-1))^N e^{(x-1)f_1(t)} 
\,,\label{eq:ansatz1}
\end{equation}
which results in the following 2D ODEs
\begin{eqnarray}
\dot{f}_1 &=& g-k f_1 \nonumber \\
\dot{f}_2 &=& \lambda(1-f_2)-\mu f_1 f_2 
\,.\label{eq:ans1ode}
\end{eqnarray}
These equations have a particularly simple physical explanation - they
correspond to the deterministic chemical kinetics equations since $f_1$ and
$N*f_2$ are equal to the average numbers of $R^*$ and $A$, respectively. But
now, we may obtain probability distributions through
Eq.~(\ref{eq:ansatz1}). For example, the variance of $A$ can be easily
calculated as $\sigma^2=\langle n^2\rangle-\langle n
\rangle^2=f_2(t)-f_2^2(t)$.

\begin{figure}[h] 
\centering
\subfigure[$P(N_{A^*})$ at $t=30$]{
\includegraphics[width=8.cm]{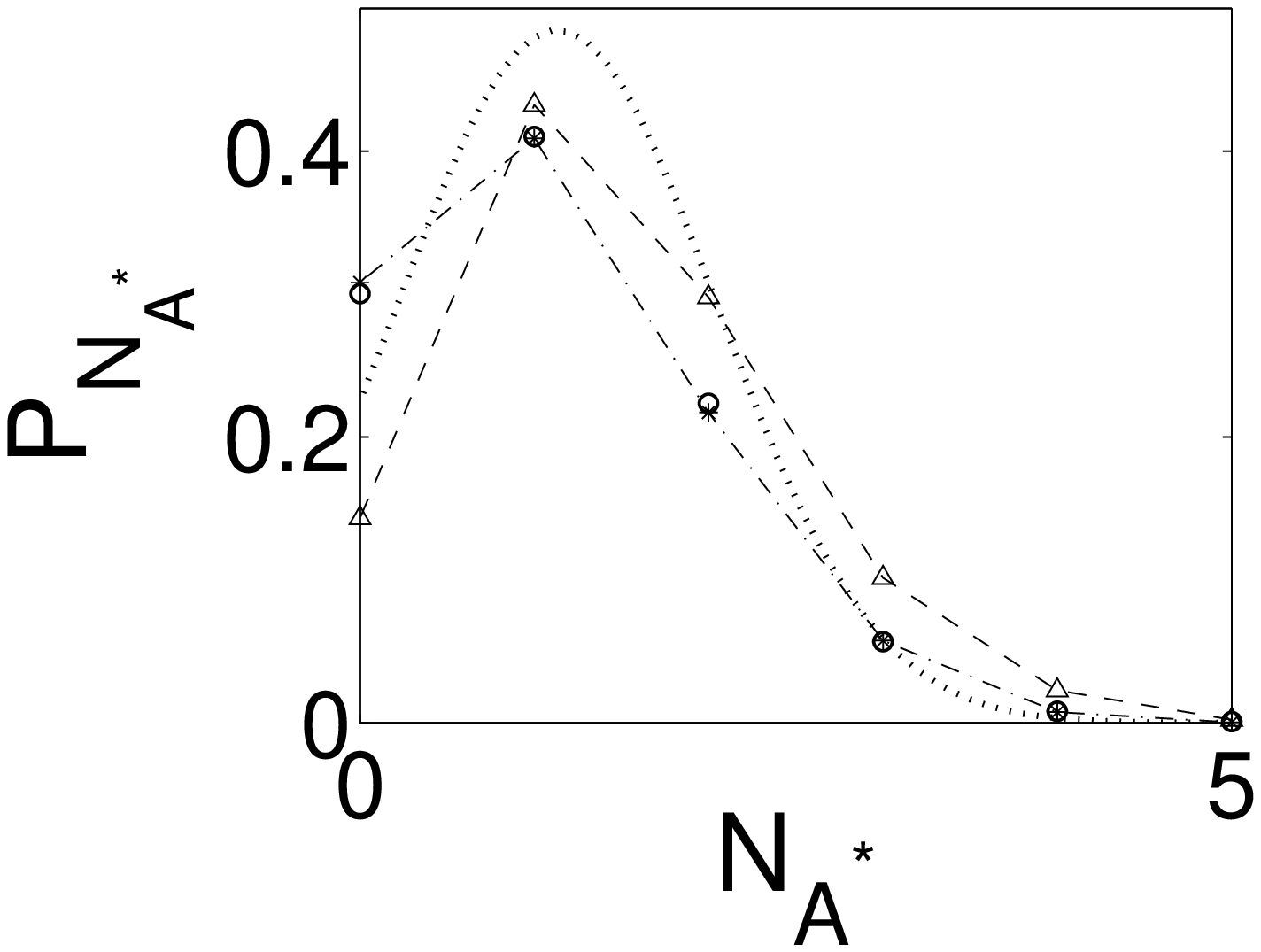}}
\subfigure[$P(N_{A^*})$ at $t=30$]{
\includegraphics[width=8.cm]{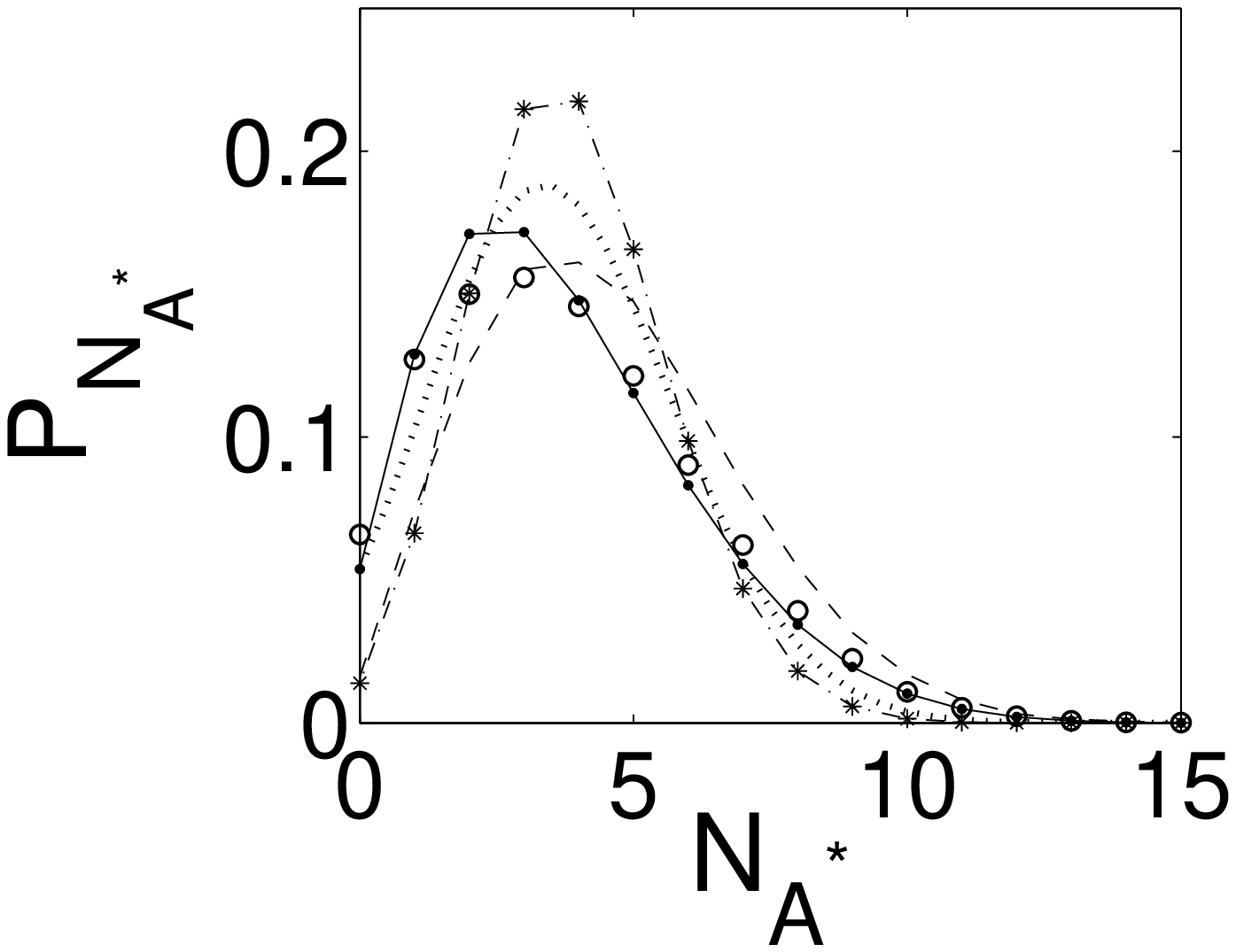}}
\caption{Comparison of the computed distributions for $N_{A^*}$ at $t=30$ for
  the \textit{2-step} cascade: Gillespie simulation (circles), one-term basis
  (dashdotted line), integral form basis (solid line), $\Omega$-expansion
  (dotted line), Langevin equation (dashed line).  (a)
  $g=10\,,k=5\,,\mu=0.02\,,\lambda=0.15$ with initial condition
  $(N_R,N_{R^*},N_A,N_{A^*})=(20,0,5,0)$ and (b)
  $g=0.2\,,k=0.1\,,\mu=0.02\,,\lambda=0.15$, with
  $(N_R,N_{R^*},N_A,N_{A^*})=(20,0,20,0)$.  }
\label{f:cas2dln1}
\end{figure}

These ODEs can be solved exactly and we show in Fig. \ref{f:cas2dln1} the
probability distribution of $A^*$ at $t=30$ for two sets of parameter
values. Also shown in the figure are results obtained from calculations using
more traditional techniques. The first set of reaction rate parameters were
chosen as $g=10\,,k=5\,,\mu=0.02\,,\lambda=0.15$, with the initial conditions
$(N_R,N_{R^*},N_A,N_{A^*})=(20,0,5,0)$. Since the $R-R^*$ reaction is much
faster than the $A-A^*$ reaction, one expects Eq.~(\ref{eq:phi0th1}) to be a
good approximation~\cite{lan1cas}. Indeed, in Fig. \ref{f:cas2dln1}(a), the
variational ansatz Eq.~(\ref{eq:ansatz1}) leads to a result that overlaps
significantly better with the exact Gillespie calculation, compared with the
results from $\Omega$-expansion and Langevin equation . The $\Omega$-expansion
result turns out to be more concentrated than the exact result, while the
Langevin equation does not work well near the left boundary, shifting the
average to the right.

For other parameter values, as long as the $R-R^*$ reaction is fast, the
ansatz Eq.~(\ref{eq:ansatz1}) works fine as expected~\cite{lan1cas}.  However,
if the first reaction is considerably slower than the second one, this ansatz becomes less
useful, as shown in Fig. \ref{f:cas2dln1}(b) for
$g=0.2\,,k=0.1\,,\mu=0.02\,,\lambda=0.15$, with
$(N_R,N_{R^*},N_A,N_{A^*})=(20,0,20,0)$. The variational result gives a too
narrow distribution. The Langevin equation is still not accurate on the left
boundary, the average being shifted to the right.

\begin{figure}[h] 
\centering
\subfigure[The average]{
\includegraphics[width=8.cm]{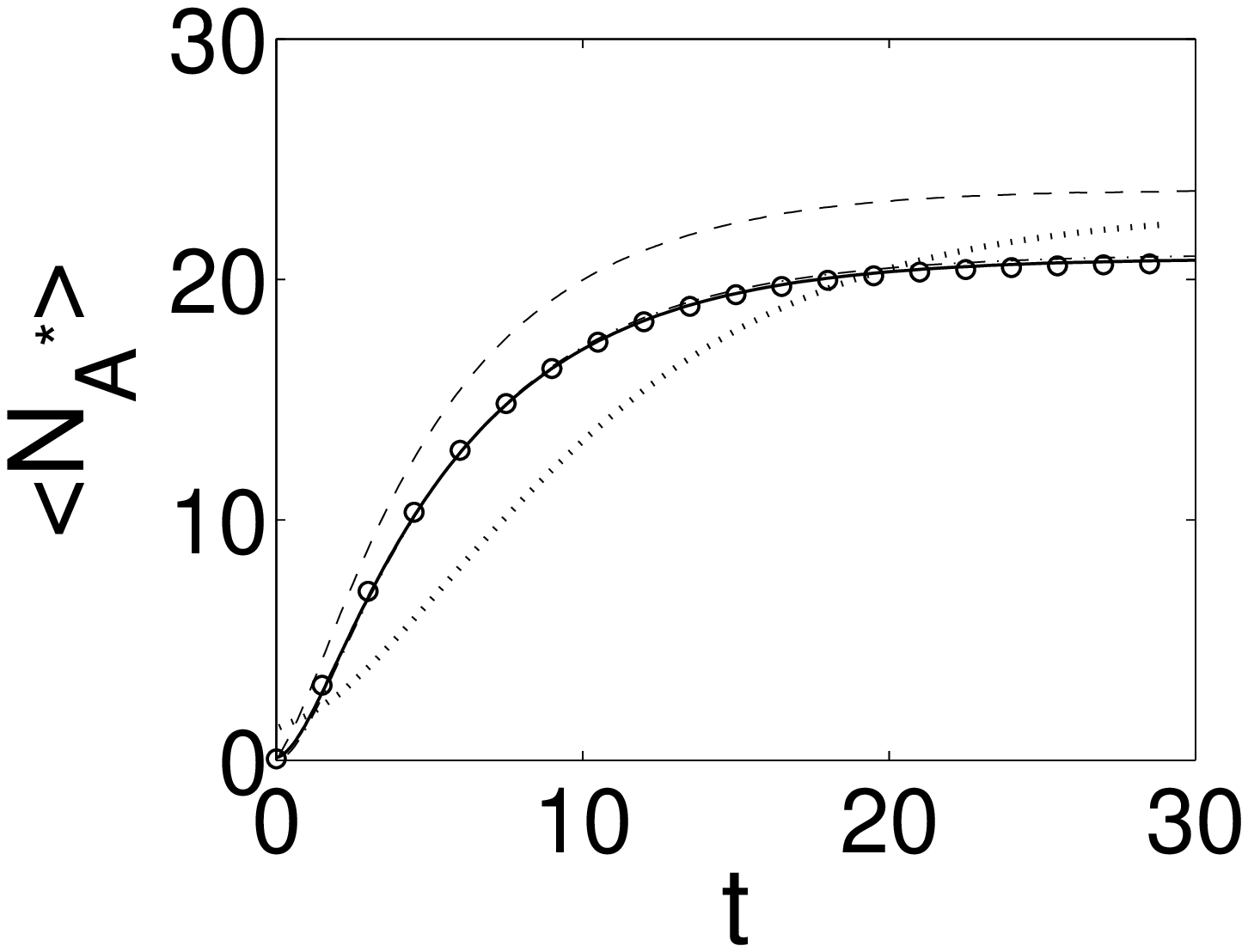}}
\subfigure[The variance]{
\includegraphics[width=8.cm]{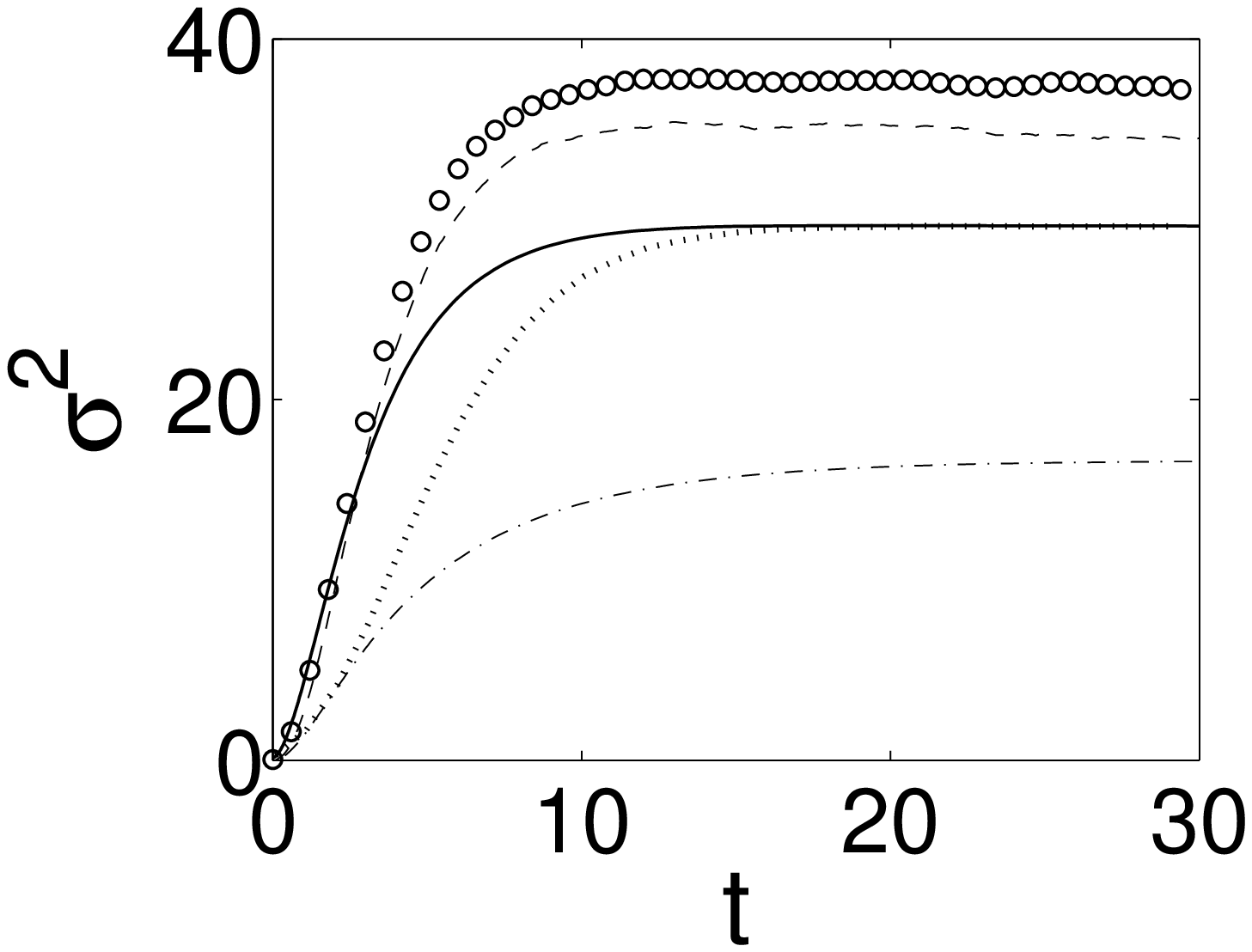}}
\caption{Comparison of the $N_{A^*}$ average and variance computed for the
  \textit{2-step} cascade in time interval $t=[0,30]$: Gillespie simulation
  (circles), one-term basis (dashdotted line), integral form basis (solid
  line), $\Omega$-expansion (dotted line) and Langevin equation (dashed line).
  $g=2\,,k=1\,,\mu=0.02\,,\lambda=0.15$ with initial condition
  $(N_R,N_{R^*},N_A,N_{A^*})=(20,0,100,0)$.}
\label{f:cas2m12ln3}
\end{figure}

In general, the ansatz (\ref{eq:ansatz1}) tends to generate a distribution
narrower than the exact one, which is also shown in Fig.
\ref{f:cas2m12ln3}(b). This can be explained as follows. The ansatz
(\ref{eq:ansatz1}) is a product of functions of $x$ and $y$ and hence only the
average particle number $f_1$ appears in the second equation of
(\ref{eq:ans1ode}). Therefore, the fluctuation generated in the $R-R^*$
reaction is absent in the treatment of the $A-A^*$ reaction. Physically, if
the first reaction is fast, then the second reaction only ``sees'' an average
number of $R^*$, with its fluctuation averaged out, and the ansatz
(\ref{eq:ansatz1}) produces accurate results (Fig.  \ref{f:cas2dln1}(a)). If
the first reaction is slow, however, then the fluctuations in the number of
$R^*$ strongly influences the $A-A^*$ reaction and the mere average $f_1$ is
not capable of passing this information. The distribution computed from ansatz
(\ref{eq:ansatz1}) only accounts for the internal fluctuation of $A-A^*$
reaction and hence has a narrower profile than the exact result. On the other
hand, despite the apparent simplicity, this ansatz allows one to estimate
fluctuations in a reaction network in a semiquantitative way, with an
extremely low computational cost, similar to solving the ordinary
deterministic kinetics equations.

\begin{figure}[h] 
\centering
\subfigure[$P(N_{A^*})$ at $t=6$]{
\includegraphics[width=8cm]{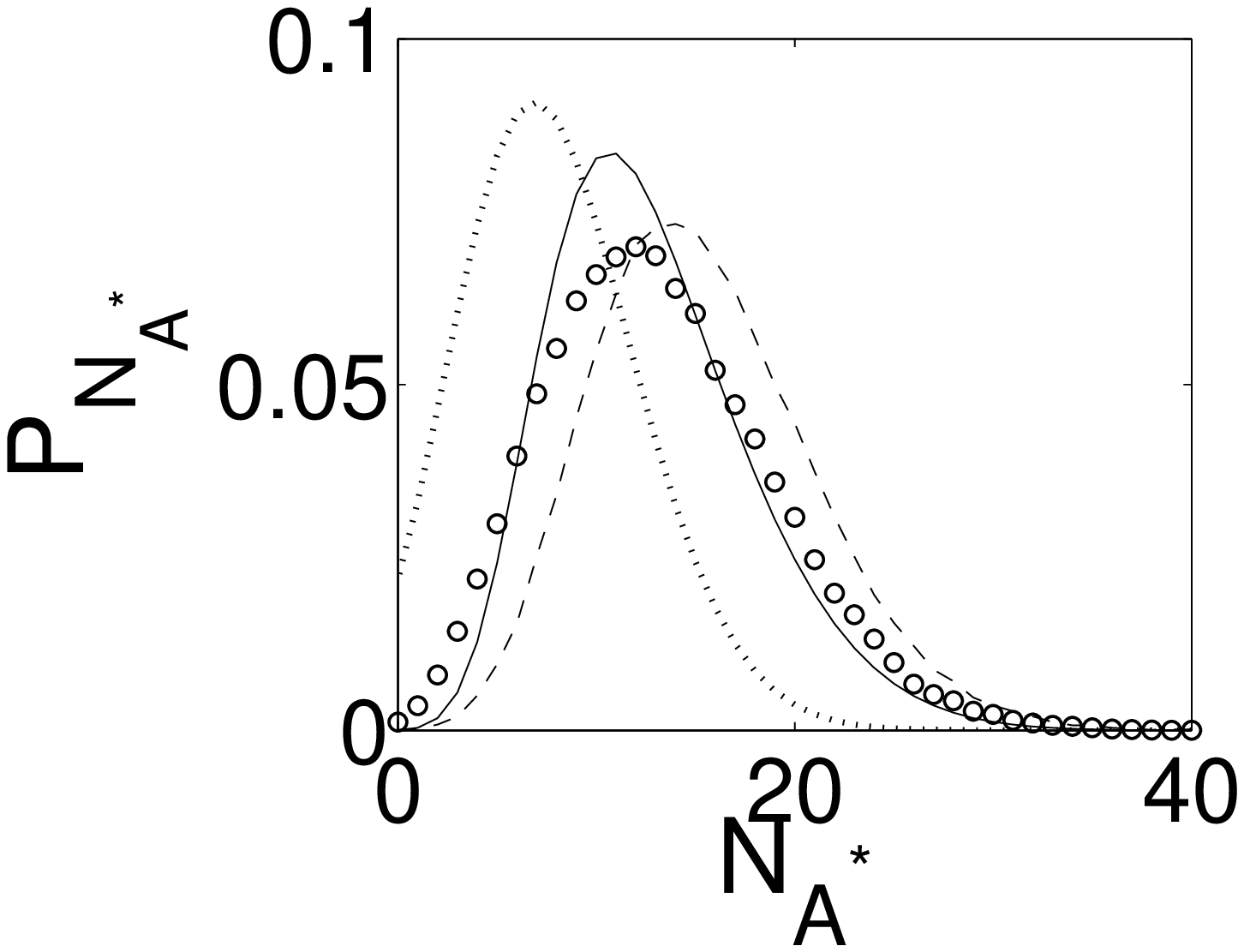}}
\subfigure[$P(N_{A^*})$ at $t=30$]{
\includegraphics[width=8.cm]{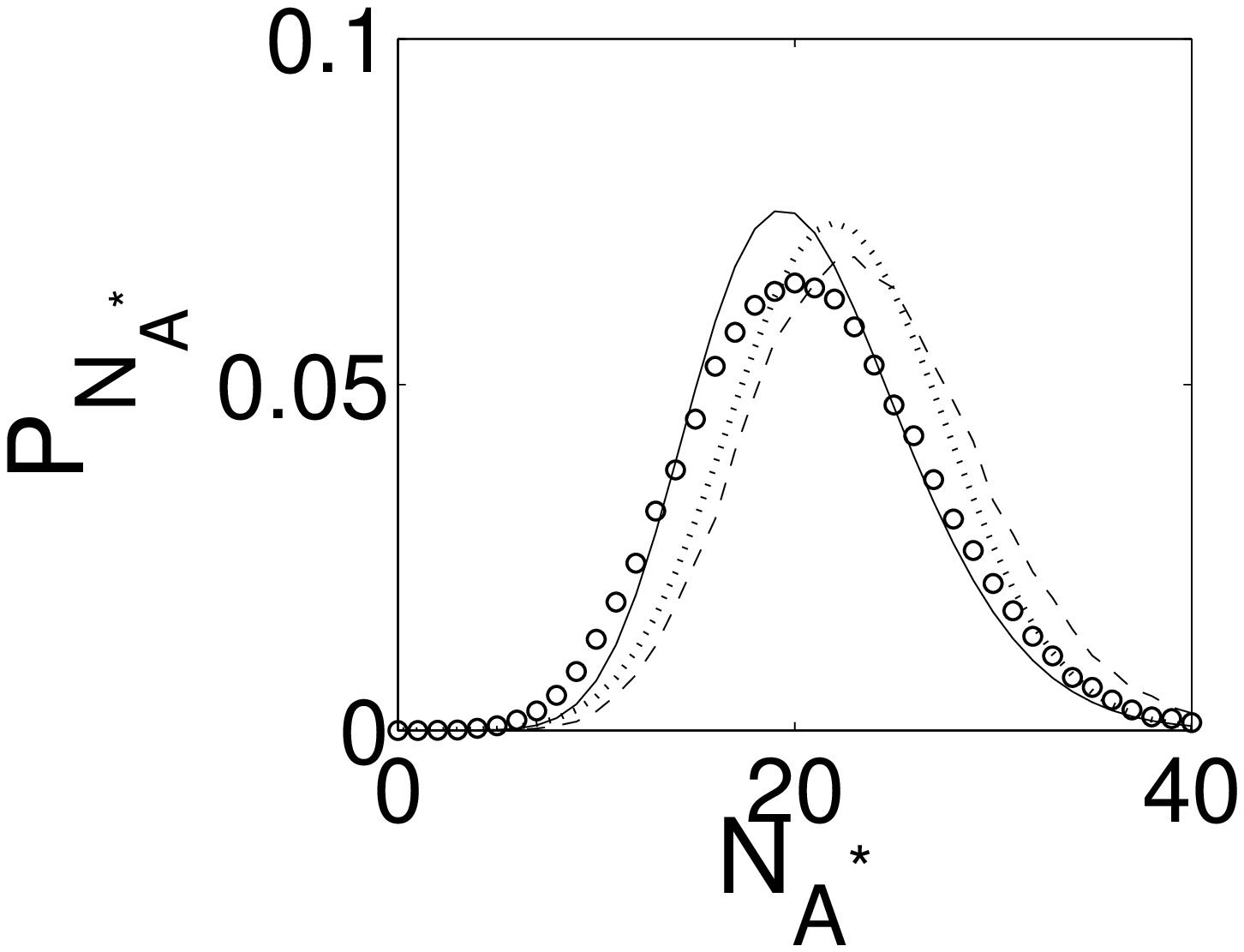}}
\caption{Comparison of the computed distributions for $N_{A^*}$ at $t=6$ and
  $t=30$ for the \textit{2-step} cascade: Gillespie simulation (circles),
  integral form basis (solid line), $\Omega$-expansion (dotted line) and
  Langevin equation (dashed line).  $g=2\,,k=1\,,\mu=0.02\,,\lambda=0.15$ with
  initial condition $(N_R,N_{R^*},N_A,N_{A^*})=(20,0,100,0)$.  }
\label{f:cas2dln3}
\end{figure}

It is straightforward to generalize ansatz (\ref{eq:ansatz1}) to longer
cascades. For example, for the 3-step cascades considered next, we may write
the right ansatz as
\begin{equation}
\Phi_R=(1+f_3(t)(z-1))^{N_2}(1+f_2(t)(y-1))^{N}e^{f_1(t)(x-1)}
\,. \label{eq:ansatz2}
\end{equation}
The resulting ODEs for $f_i(t)$'s are similar to Eq.~(\ref{eq:ans1ode}) and
have a physical interpretation related to the chemical kinetics equations, as
discussed above.

To get more accurate result, we have to convolute the number fluctuation of
$R^*$ with the number fluctuation of $A$. Since the ansatz based on simple
separation of variables does not work, we need an equation in which $x\,,y$ are
explicitly entangled. To facilitate the computation, we use the following
integral form representation:

\begin{equation}
\Phi_R(x,y) =\int_{-\infty}^{\infty}ds \frac{e^{-s^2}}{\sqrt{\pi}}
\left(1+f_2(t)e^{-(s-f_3(t))^2}(y-1)+f_1(t)(x-1) \right)^N
\,, \label{eq:ans2}
\end{equation}
where $f_1(t)$ is related to the $R-R^*$ reaction and $f_2(t)\,,f_3(t)$ are
related to the $A-A^*$ reaction. Note that $\Phi_R(1,1)=1$. For $y=1$, we get
the expected generating function for $R^*$
\begin{equation}
\Phi_R(x,1)=\left(1+f_1(t)(x-1) \right)^N \approx e^{Nf_1(t)(x-1)}
\,, \label{eq:ans2x}
\end{equation}
the above approximation being valid when $f_1(t)$ is small, which is true in
all simulations below. We could have used
\[
\frac{e^{-s^2}}{\sqrt{\pi}}\left(1+f_2(t)e^{-(s-f_3(t))^2}(y-1)\right)^N \exp((s-f_1(t))^2(x-1))
\]
in the integrand of (\ref{eq:ans2}) to achieve a larger range of $f_1$. But
when the number of $R^*$ is small, ansatz (\ref{eq:ans2}) produces better
results, probably due to its more convoluted form.

Now we can control both the average and the variance of $A$ by manipulating
$f_2(t)$ and $f_3(t)$. Roughly speaking, $f_2(t)$ controls the average and
$f_3(t)$ controls the variance. For the same parameter set shown in
Fig. \ref{f:cas2dln1}(b), we did the computation by using ansatz
Eq.~(\ref{eq:ans2}) and displayed the result in the same figure (solid line).
It matches closely with the exact result, better than all other computations.

To show the effectiveness of the ansatz (\ref{eq:ans2}), we use it to do one
more computation with $g=2\,,k=1\,,\mu=0.02\,,\lambda=0.15$ and the initial
condition $(N_R,N_{R^*},N_A,N_{A^*})=(20,0,100,0)$. In Fig. \ref{f:cas2dln3},
the distributions of $A^*$ are displayed at $t=6$ and $t=30$. Although, the
result from ansatz (\ref{eq:ans2}) is slightly narrower than the Gillespie
computation, they match very well both at $t=6$ and at $t=30$.  Actually, this
is true for all times as can be seen in Fig. \ref{f:cas2m12ln3} where the time
evolution of the average and the variance are depicted. In
Fig. \ref{f:cas2dln3} and \ref{f:cas2m12ln3}, the average of $A^*$ from
Langevin equation is always greater than the exact result as explained before,
although the variance is computed accurately. Curiously, the average from
$\Omega$-expansion is smaller initially but later grows larger than the exact
one. We also plotted the computation results from Eq.~(\ref{eq:ansatz1}). It
gives a smaller variance even though the average is quite accurate. In this
case, the $R^*$ fluctuation is important.

\subsection{Application to a three-step amplification cascade}

It is not hard to write ans{\" a}tze similar to Eq.~(\ref{eq:ans2}) for longer or
more complicated cascades. In this section, we demonstrate the use of the
variational method for a 3-step cascade with and without feedback loop. In the
next section, we will write the equation for a 4-step cascade.

\begin{figure}[h] 
\centering
\subfigure[without feedback]{
\includegraphics[width=8.cm]{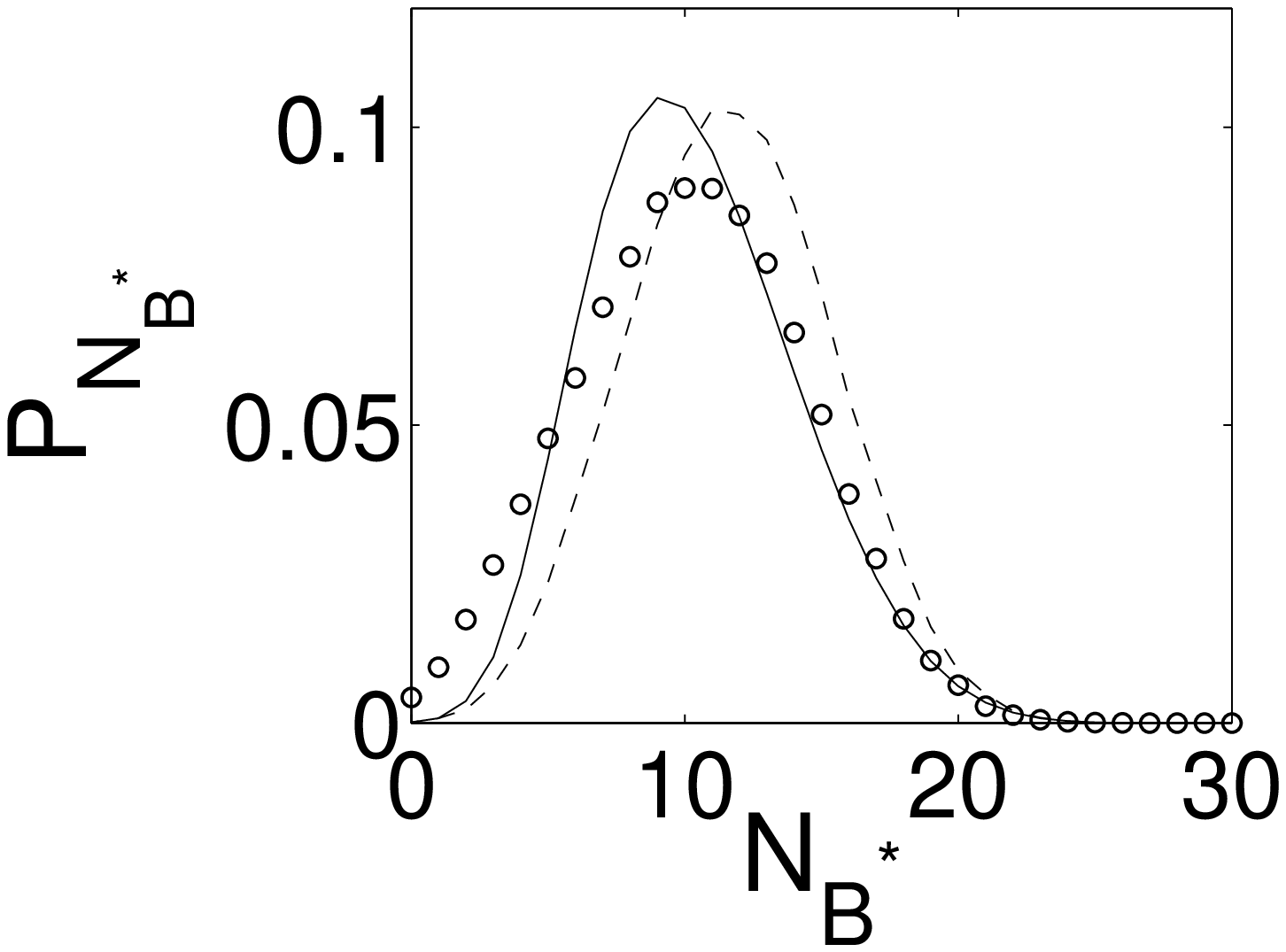}}
\subfigure[with feedback]{
\includegraphics[width=8.cm]{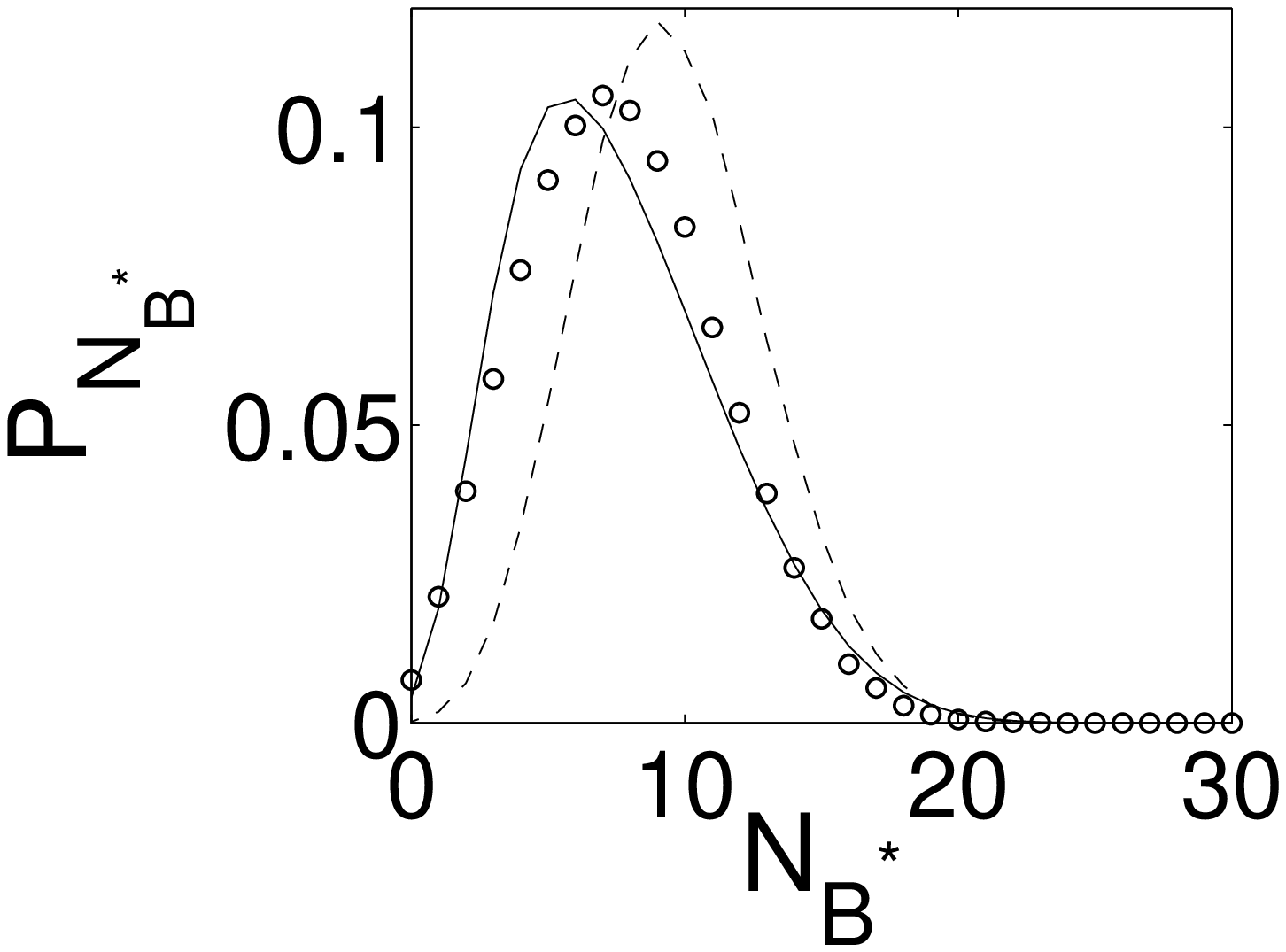}}
\caption{Comparison of the computed distributions for $N_{B^*}$ at $t=60$ for
  the \textit{3-step} cascade without (a) and with (b) negative feedback:
  Gillespie simulation (circles), integral form basis (solid line), Langevin
  equation (dashed line).
  $g=0.2\,,k=0.1\,,\mu=0.02\,,\lambda=0.15\,,\mu_2=0.01\,,\lambda_2=0.07\,,
  \mu_3=0.01$ with initial conditions 
  $(N_R,N_{R^*},N_A,N_{A^*},N_B,N_{B^*})=(20,0,20,0,30,0)$.}
\label{f:cas3dln1}
\end{figure}

\begin{figure}[h]
\centering
\subfigure[Without feedback]{
\includegraphics[width=8.cm]{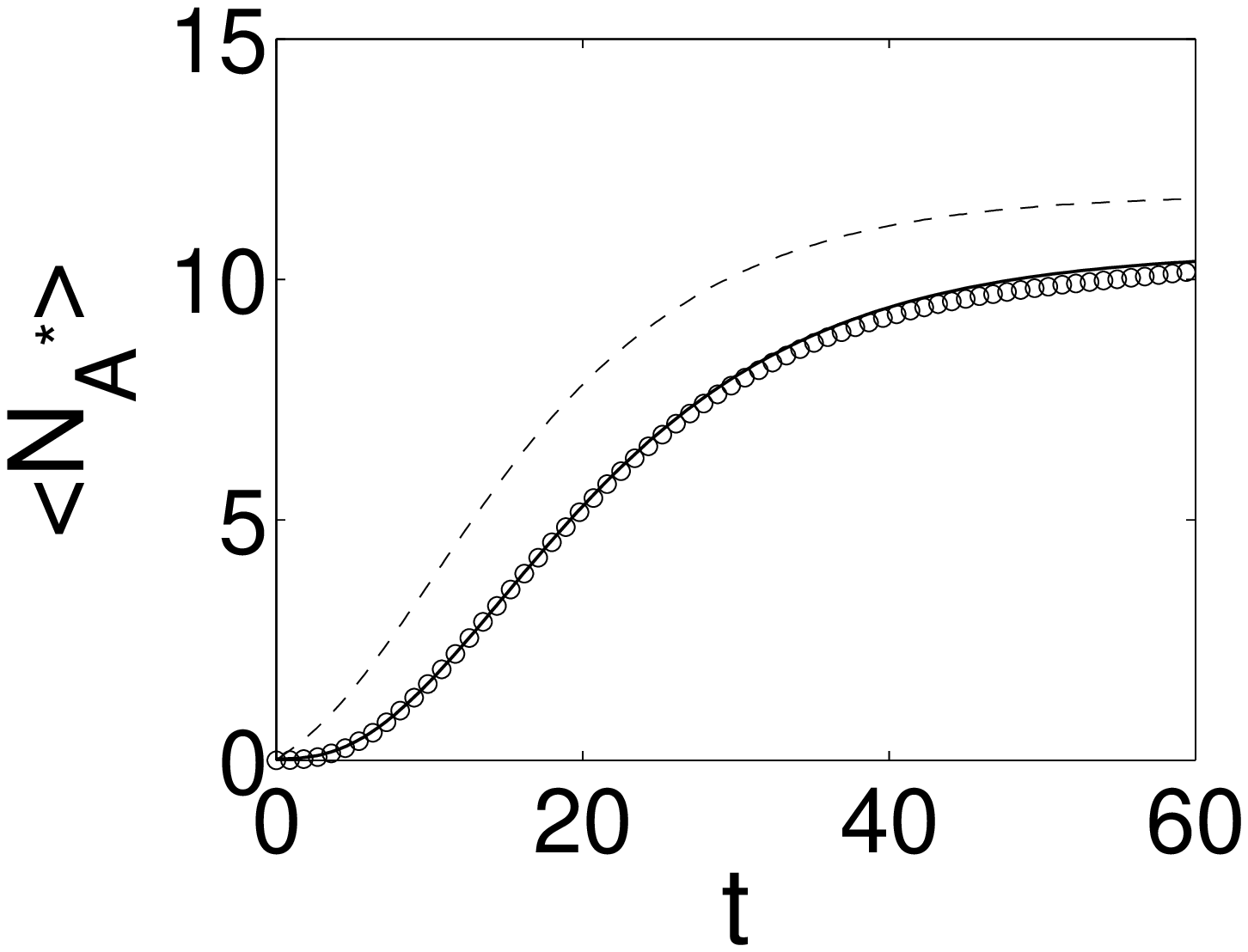}}
\subfigure[With feedback]{
\includegraphics[width=8.cm]{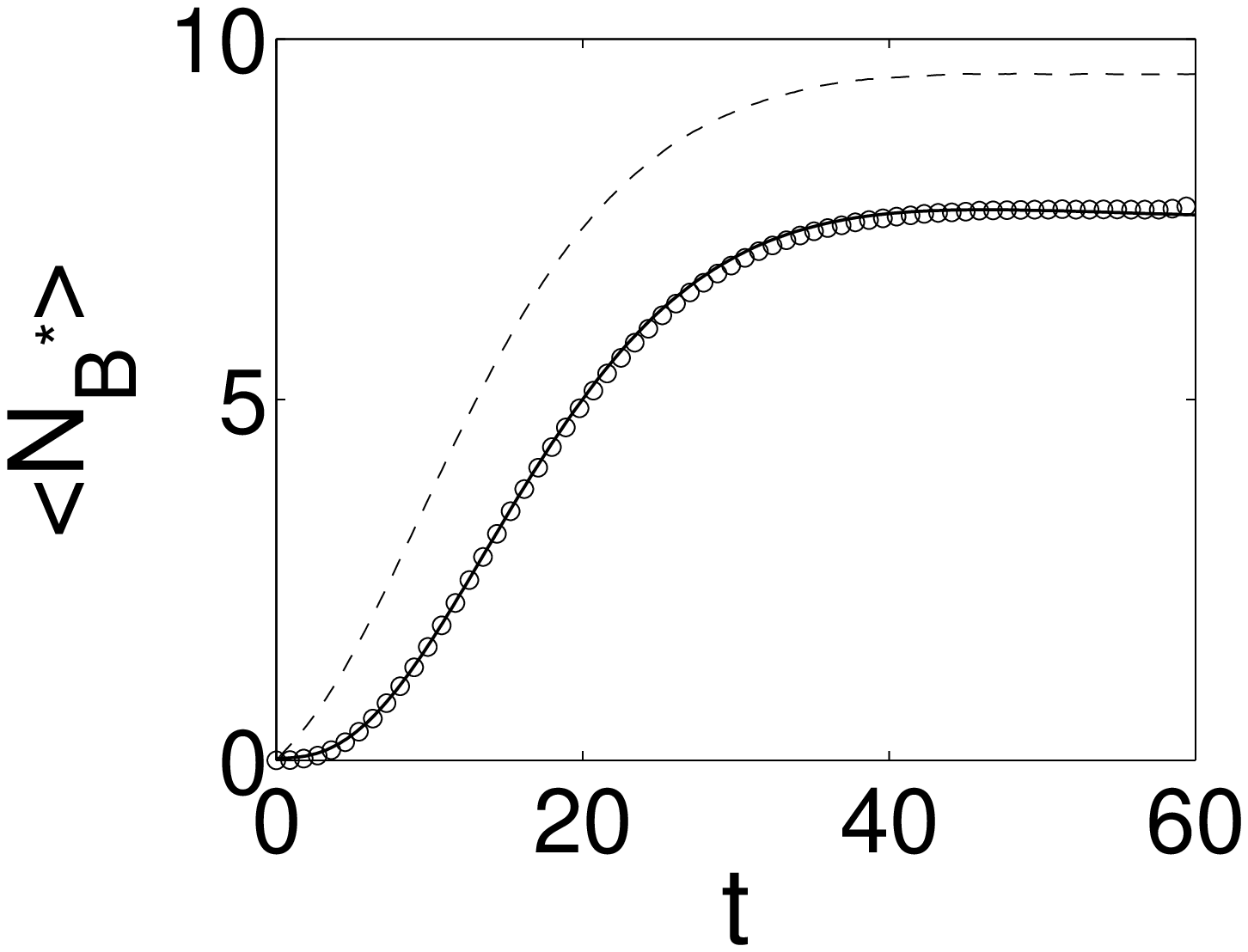}}
\caption{ Comparison of the $N_{B^*}$ average computed for the \textit{3-step}
  cascade without (a) and with (b) negative feedback in time interval
  $t=[0,60]$: Gillespie simulation (circles), integral form basis (solid line),
  Langevin equation (dashed line).
  $g=0.2\,,k=0.1\,,\mu=0.02\,,\lambda=0.15\,,\mu_2=0.01\,,\lambda_2=0.07\,,
  \mu_3=0.01$ with initial condition
  $(N_R,N_{R^*},N_A,N_{A^*},N_B,N_{B^*})=(20,0,20,0,30,0)$.}
\label{f:cas3m1ln1}
\end{figure}

\begin{figure}[h] 
\centering
\subfigure[Without feedback]{
\includegraphics[width=8.cm]{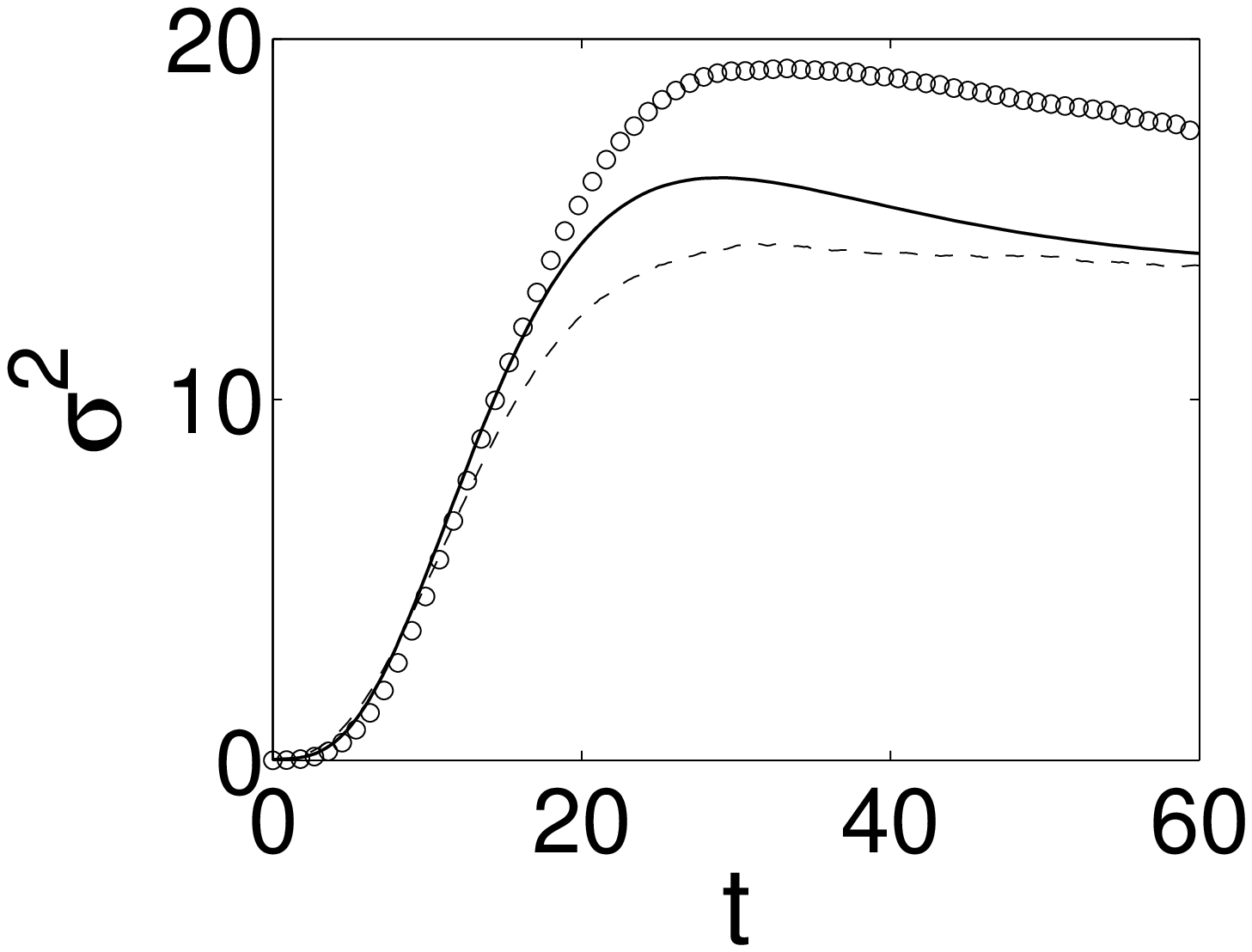}}
\subfigure[With feedback]{
\includegraphics[width=8.cm]{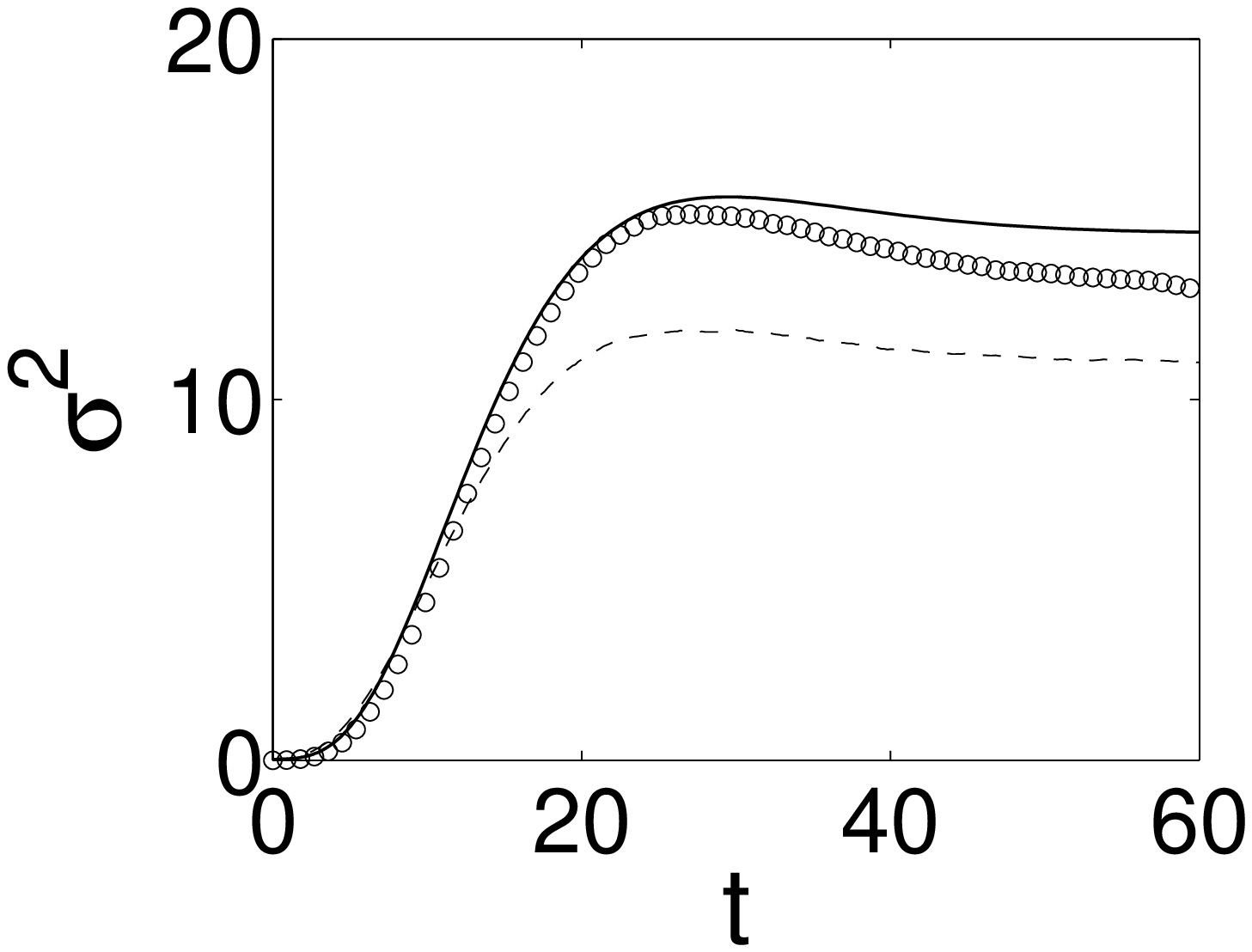}}
\caption{ Comparison of the $N_{B^*}$ variance computed for the
  \textit{3-step} cascade without (a) and with (b) negative feedback in time
  interval $t=[0,60]$: Gillespie simulation (circles), integral form basis
  (solid line), Langevin equation (dashed line).
  $g=0.2\,,k=0.1\,,\mu=0.02\,,\lambda=0.15\,,\mu_2=0.01\,,\lambda_2=0.07\,,
  \mu_3=0.01$ with initial condition
  $(N_R,N_{R^*},N_A,N_{A^*},N_B,N_{B^*})=(20,0,20,0,30,0)$.}
\label{f:cas3m2ln1}
\end{figure}

Assume that $A^*$ catalyzes a subsequent enzyme activation/deactivation
reaction $B\rightleftharpoons B^*$ with a forward rate $\mu_2$ and a backward
decay rate $\lambda_2$. The total number $N_2$ of $B$ and $B^*$ is a constant
during the reaction. Following similar procedures as before, we found that the
generating function $\Psi(x,y,z)$ satisfies
\begin{eqnarray}
\frac{\partial \Psi}{\partial t} &=& (1-z)(-\mu_2 y\frac{\partial^2}{\partial y\partial z}
-\lambda_2 N_2+(\lambda_2 z+\mu_2*N)\frac{\partial}{\partial z})\Psi \nonumber \\
&+& (1-y)(\mu x\frac{\partial^2}{\partial
  x \partial y}-\lambda N+\lambda y \frac{\partial}{\partial y})\Psi
  +g(x-1)\Psi-k(x-1)\frac{\partial \Psi}{\partial x} 
\,, \label{eq:cas3d}
\end{eqnarray}
where the first term describes the $B-B^*$ reaction.
The ansatz similar to Eq.~(\ref{eq:ans2}) reads
\begin{equation}
\Phi_R(x,y,z) =\int_{-\infty}^{\infty}ds \frac{e^{-s^2}}{\sqrt{\pi}}
\left(1+f_4(t)e^{-(s-f_5(t))^2}(z-1)\right)^{N_2}
\left(1+f_2(t)e^{-(s-f_3(t))^2}(y-1)+f_1(t)(x-1) \right)^N
\,, \label{eq:ans3}
\end{equation}

where $f_4(t),f_5(t)$ describes the $B-B^*$ reaction. The calculation results
from this ansatz are shown in Fig. \ref{f:cas3dln1}(a), \ref{f:cas3m1ln1}(a)
and \ref{f:cas3m2ln1}(a).

Shown in Fig. \ref{f:cas3dln1}(a) is the $B^*$ distribution computed from
different methods. Ansatz (\ref{eq:ans3}) computation matches very well with
the exact solution while the Langevin profile is shifted to the right. On the
left boundary, both ansatz (\ref{eq:ans3}) and Langevin equation approach zero
while the exact solution has a finite value there. Interestingly, the variance
shows a maximum value during the evolution as displayed in
Fig. \ref{f:cas3m2ln1}(a). The computation from ansatz (\ref{eq:ans3})
captures this non-monotonous behavior accurately which is not obvious at all
in the Langevin computation.

\begin{figure}[h]
\includegraphics[height=6cm]{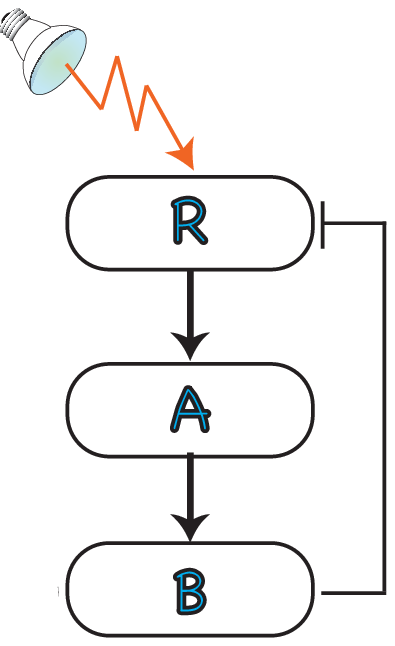}
\caption{An inactive receptor $R$, when activated by a signal, activates
  downstream protein $A$, which in turn activates protein $B$. In a negative
  feedback loop, $B$ downregulates the $R$ activation.}
\noindent 
\label{f:3step-react}
\end{figure}

Next, we consider a 3-step signaling cascade with a feedback loop. For
example, we can imagine a reaction in which $B^*$ turns off the $R^*$
signaling, by catalyzing the $R^* \to R$ decay at a rate $\mu_3$
(Fig. \ref{f:3step-react}). Mathematically, this corresponds to adding an
extra term
\[
\mu_3(1-y)(N_2\partial \Psi/\partial x-z \partial^2 \Psi/\partial x \partial
z)
\]
to the right hand side of PDE (\ref{eq:cas3d}). We may still use the same
right ansatz (\ref{eq:ans3}) and the results are displayed in Fig.
\ref{f:cas3dln1}(b), \ref{f:cas3m1ln1}(b) and \ref{f:cas3m2ln1}(b).
Surprisingly, despite the time scale mixing and nonlinearity, the variational
computation matches even better with the exact result than without feedback
(compare \ref{f:cas3dln1} (a) and (b)). The relative shift of the average
computed from the Langevin equation increases. The maximum in the variance
still exists but its height decreases with the variance itself. In this case,
it seems that the negative feedback sharpens the signal.

\subsection{Application to a four-step amplification cascade}

\begin{figure}[h] 
\centering

\includegraphics[width=8.cm]{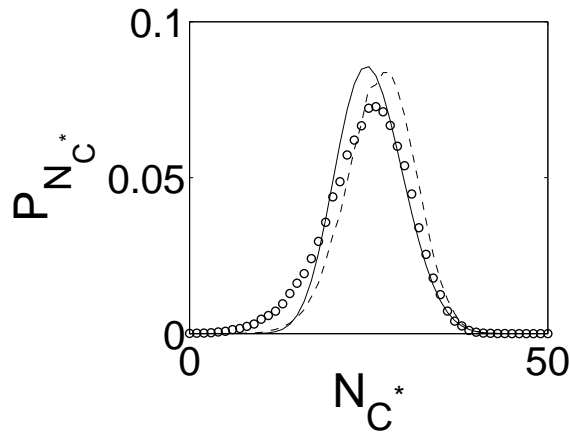}
\caption{ Comparison of the computed distributions for $N_{C^*}$ at $t=100$
  for the \textit{4-step} cascade: Gillespie simulation (circles), integral
  form basis (solid line) and Langevin equation (dashed line).
  $g=0.2\,,k=0.1\,,\mu=0.02\,,\lambda=0.15\,,\mu_2=0.01\,,\lambda_2=0.07\,,\mu_3=0.005\,,\lambda_3=0.05$
  with initial condition
  $(N_R,N_{R^*},N_A,N_{A^*},N_B,N_{B^*},N_C,N_{C^*})=(20,0,20,0,30,0,50,0)$.}
\label{f:cas4dln1}
\end{figure}

\begin{figure}[h] 
\centering
\subfigure[The average]{
\includegraphics[width=8.cm]{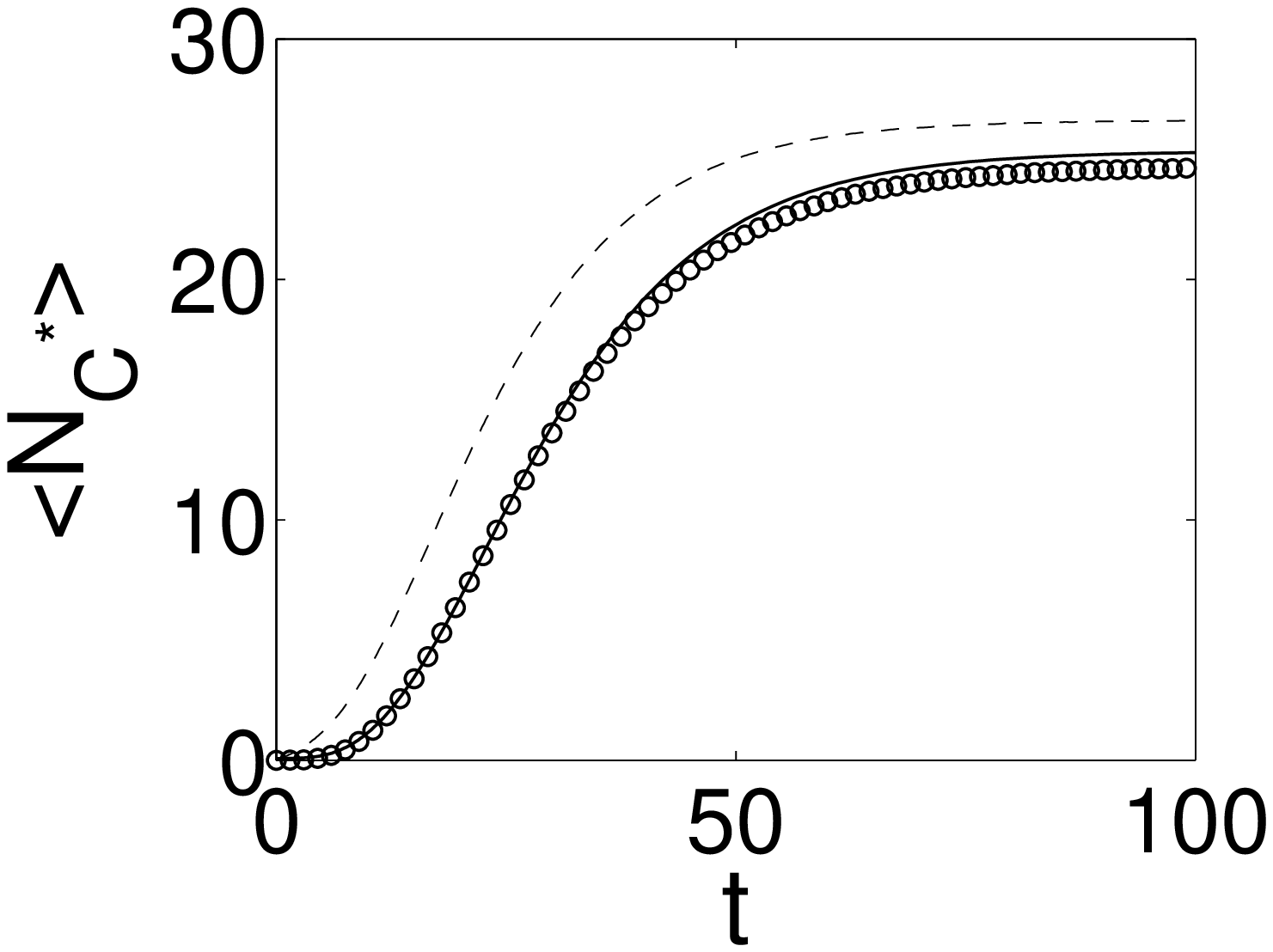}}
\subfigure[The variance]{
\includegraphics[width=8.cm]{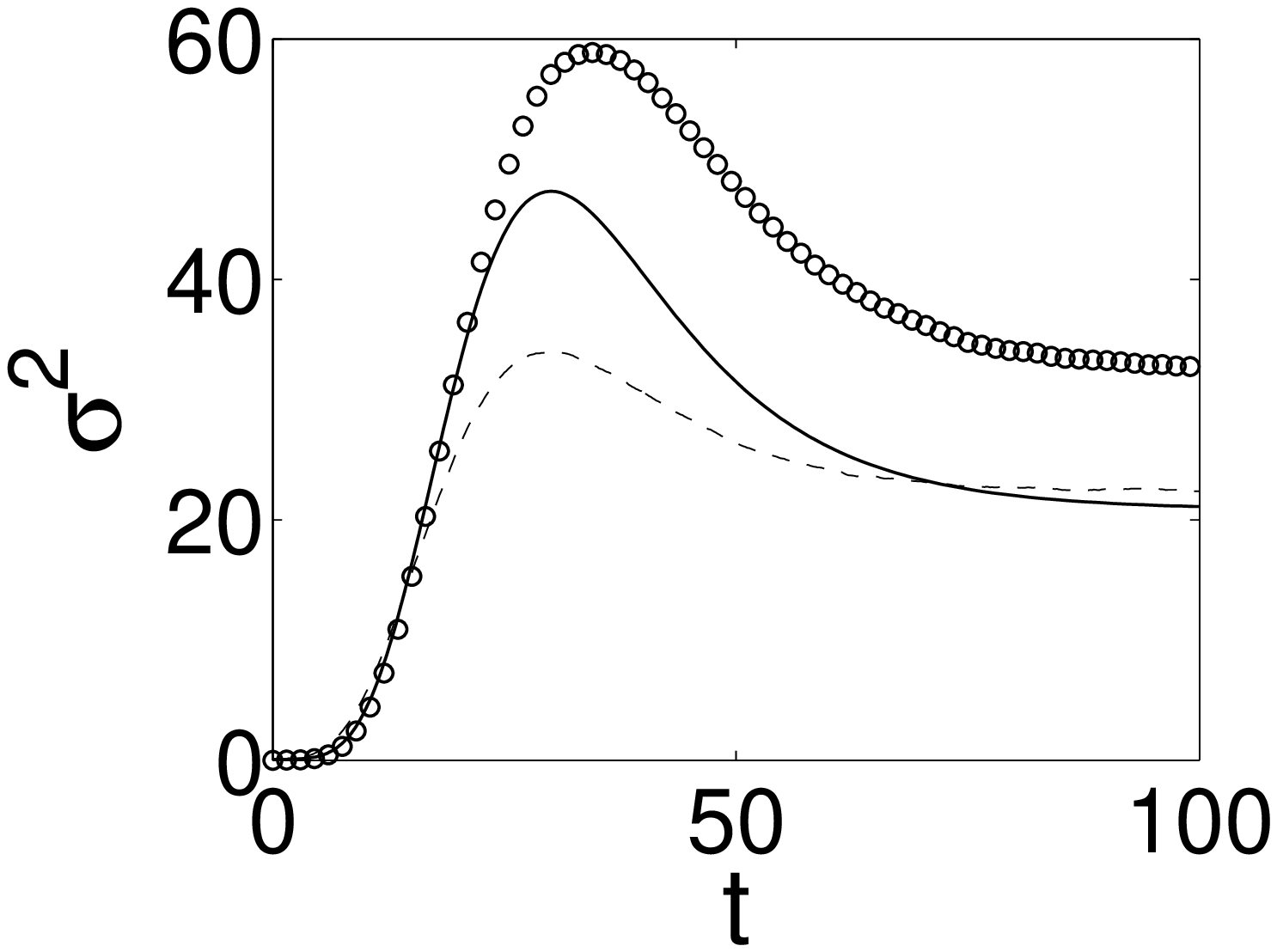}}
\caption{ Comparison of the $N_{C^*}$ average and variance computed for the
  \textit{4-step} cascade in time interval $t=[0,100]$: Gillespie simulation
  (circles), integral form basis (solid line) and Langevin equation (dashed
  line).
  $g=0.2\,,k=0.1\,,\mu=0.02\,,\lambda=0.15\,,\mu_2=0.01\,,\lambda_2=0.07\,,\mu_3=0.005\,,\lambda_3=0.05$
  with initial condition
  $(N_R,N_{R^*},N_A,N_{A^*},N_B,N_{B^*},N_C,N_{C^*})=(20,0,20,0,30,0,50,0)$.}
\label{f:cas4m12ln1}
\end{figure}

Our last demonstration of the variational method is concerned with a 4-step
cascade. We append a further enzymatic reaction $C \rightleftharpoons C^*$ to
our 3-step cascade without feedback. In this reaction, the protein $C$ is
switched on with a rate $\mu_3$ by $B^*$ and decays at a rate
$\lambda_3$. Again, the total number $N_3$ of $C$ and $C^*$ is a constant
during the reaction. Routinely, we add the corresponding extra term
\[
(1-w)(-\mu_3 z\frac{\partial^2}{\partial z \partial w}-\lambda_3 N_3 
+(\lambda_3 w+\mu_3 N_2)\frac{\partial}{\partial w})\Psi
\]
to the right hand side of Eq.~(\ref{eq:cas3d}). The right ansatz is also
postulated following the previous pattern,
\begin{eqnarray}
\Phi_R(x,y,z,w) &=& \int_{-\infty}^{\infty}ds \frac{e^{-s^2}}{\sqrt{\pi}}
\left(1+f_6(t)e^{-(s-f_7(t))^2}(w-1)\right)^{N_3}
\left(1+f_4(t)e^{-(s-f_5(t))^2}(z-1)\right)^{N_2} \nonumber \\
& & \left(1+f_2(t)e^{-(s-f_3(t))^2}(y-1)+f_1(t)(x-1) \right)^N
\,, \label{eq:ans4}
\end{eqnarray}
where $f_6(t),f_7(t)$ describes the $C-C^*$ reaction. The computations for a
particular set of parameters were carried out and the results are depicted in
Fig. \ref{f:cas4dln1} and \ref{f:cas4m12ln1}.

In Fig. \ref{f:cas4dln1}, the distributions of $C^*$ at $t=100$ from different
calculations agree with each other very well. The variational profile is
slightly narrower than the exact one but the averages overlap at all times
(see Fig. \ref{f:cas4m12ln1}(a)). Now, the maximum in the variance becomes
more pronounced. Even the Langevin computation clearly displays this feature
in Fig. \ref{f:cas4m12ln1}(b) though its peak is considerably smaller than the
exact one.

\section{The advantages and drawbacks of the variational principle} 

The enzymatic reaction cascades of various lengths considered in section
\ref{sect:num} are a common occurrence, for instance, in the MAPK family of
signaling cascades~\cite{fer97mech,ch96ul,bur97mek,seg95map}.  It is
straightforward to extend the use of our variational scheme to more complex
cases, to cascades or networks with complex topology. In general, the
generating function is to be postulated in an integral form, as demonstrated
earlier for the 2-, 3-, and 4-step cascades, with the time-dependent parameter
functions being determined by a set of ODEs derived from the variational
principle. Our scheme may be used to treat both the small and large particle
number systems. It is many orders of magnitude computationally more efficient
in computing the distributions compared with alternative numerical simulation
techniques such as the Gillespie algorithm or Langevin equation.

However, in the current form, the ansatz has a number of practical
limitations, discussed next. It is difficult to represent efficiently
distributions with multiple peaks, for example, or to directly compute
transition rates between two deterministically stable states, a common
scenario in a gene switch modeling.  Another problem is that the derived set
of ODEs is quite complicated, thus, symbolic algebra software is necessary to
carry out the necessary manipulations. The method accuracy may also depend on
the choice of the left ansatz. We chose the current left ansatz form from
several trials for simplicity and efficiency. Sometimes when
Eq.~\ref{eq:varg2}, the time evolution of the unknown functions may result in
a possible singularity, requiring workarounds. For reaction types other than
the enzymatic one discussed here, such as the binding reactions, the current
basis functions may not work properly, since the total particle number of one
species (including both the activated and inactive ones) is required to be
constant. This may not be true for some arbitrary reaction, necessitating
development of new basis functions. However, this is straightforward, and the
general principles and considerations that were discussed are expected to
apply to those cases as well.

In general, sufficient accuracy may be achieved with a large number of basis
functions. The probability distributions are obtained when $\Phi_R$ is
expanded at $x=0$ and the moments are obtained when $\Phi_R$ expanded at
$x=1$. Therefore, $\Psi$ together with its derivatives is to be well
approximated in the whole interval $x \in [0,1]$. However, the variation
equations~(\ref{eq:varg1}) only consider the validity of Eq.~(\ref{eq:mg}) in
the neighborhood of $x=1$. It is difficult to estimate the error bounds of
$\Psi$ and especially its derivatives near $x=0$, though we know that it
generally decreases with increasing accuracy at $x=1$. The choice of basis
function, therefore, is essential for the variational technique to be
successful.

\begin{figure}[h] 
\centering
\subfigure[$P(N_{A^*})$ at $t=30$]{
\includegraphics[width=8.cm]{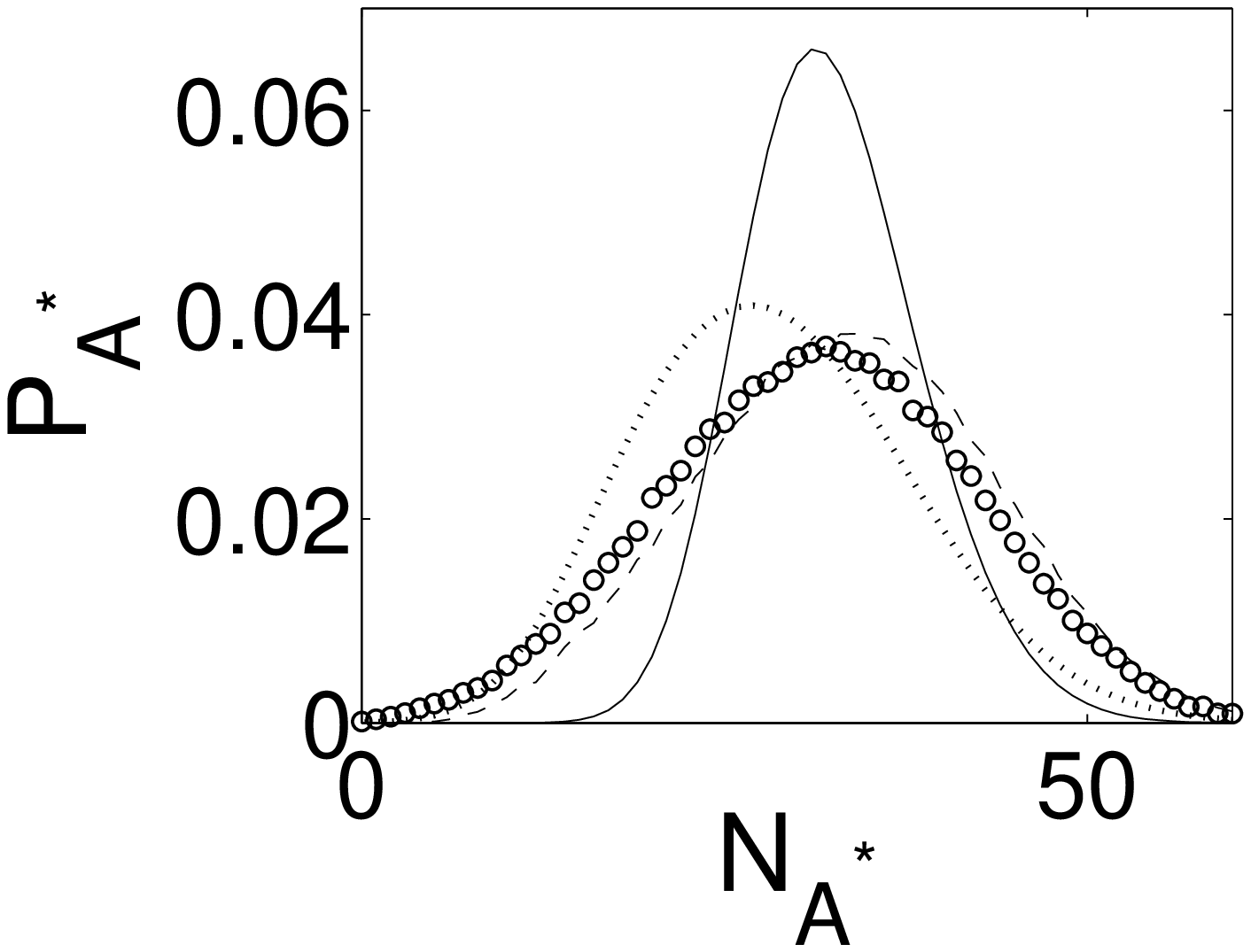}}
\subfigure[$P(N_{A^*})$ at $t=30$]{
\includegraphics[width=8.cm]{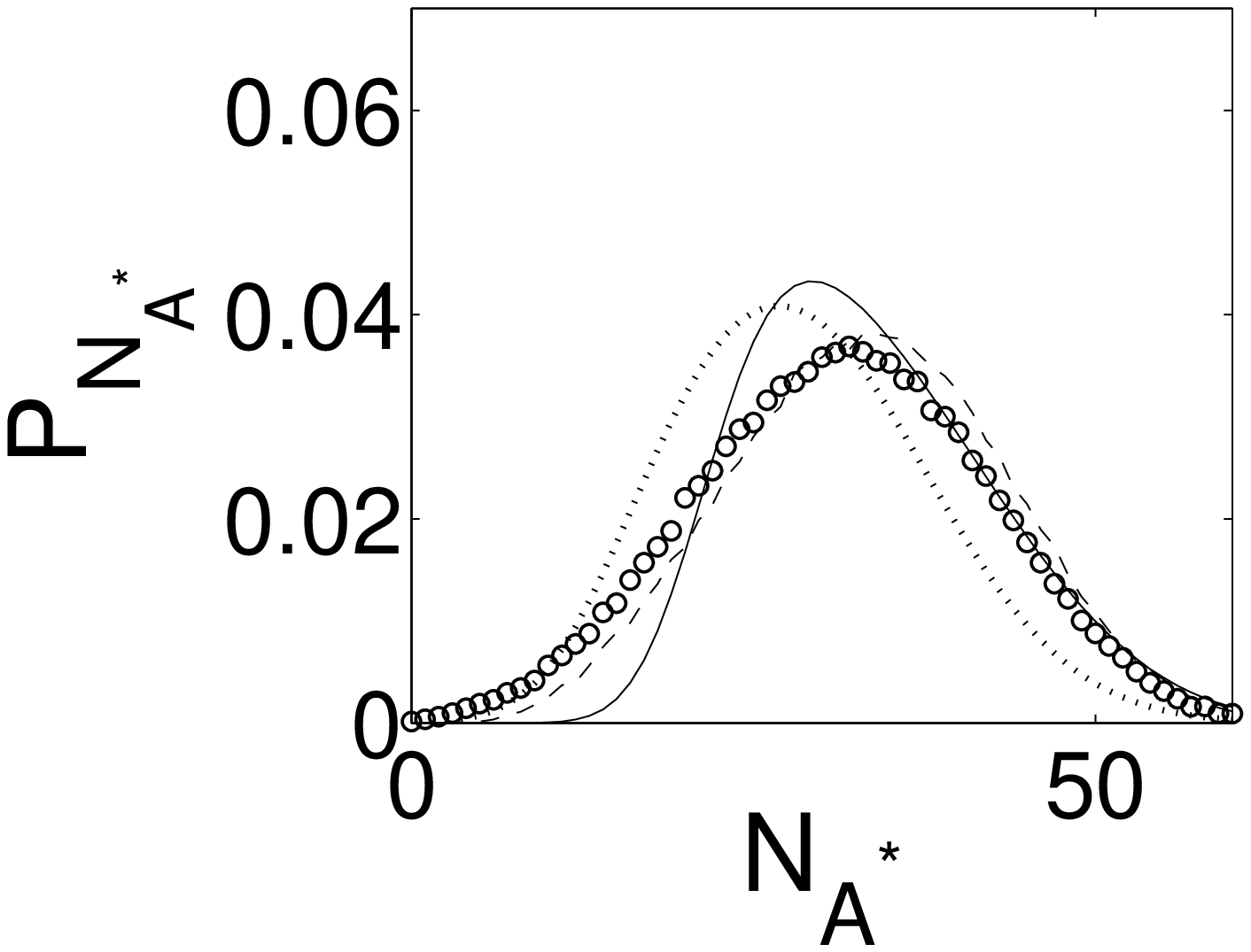}}
\caption{ Comparison of the computed distributions for $N_{A^*}$ at $t=30$ for
  the \textit{2-step} cascade: Gillespie simulation (circles), integral form
  basis (solid line), $\Omega$-expansion (dotted line) and Langevin equation
  (dashed line).  $g=0.4\,,k=0.1\,,\mu=0.02\,,\lambda=0.15$ with initial
  condition $(N_R,N_{R^*},N_A,N_{A^*})=(20,0,100,0)$.
  (a)$f_2(t=30)=0.7304\,,f_3(t=30)=0.1413$ calculated by integration of
  Eq.~\ref{eq:ans1ode}, (b)$f_2(t=30)=0.78\,,f_3(t=30)=0.3$ estimated by best
  fitting the exact solution.  Shown in the picture is only the distribution
  profile on $[0,60]$ with other part close to zero.  }
\label{f:cas2dln2}
\end{figure}

From the above considerations one may expect that the variational principle
itself may still be improved. In Fig. \ref{f:cas2dln2}, we show the
distributions of $A^*$ calculated with different methods with a parameter set
$g=0.4\,,k=0.1\,,\mu=0.02\,,\lambda=0.15$ and initial conditions
$(N_R,N_{R^*},N_A,N_{A^*})=(20,0,100,0)$. In this case, the $R-R^*$ reaction
is unusually slow compared with the $A-A^*$ reaction so that the large
fluctuations in the first reaction are retained in the second one. To obtain a
highly accurate solution in this parameter regime, a special convolution form
was used to solve the generating function PDE in our previous
work\cite{lan1cas}. The current variational scheme, however, underestimates
the $A^*$ distribution variance. By manually adjusting the $f_2\,,f_3$, we may
obtain a much better fit (solid line in Fig. \ref{f:cas2dln2}(b)),
demonstrating that the variational calculation does not necessarily provide an
optimal solution. However, these results suggest that the present
time-dependent basis sets are powerful enough to account for these extremely
broad distributions.  From the experience of numerical solution of ODEs and
the conventional variational method in quantum mechanics, a better variational
strategy may be to consider simultaneously the validity of Eq.~(\ref{eq:mg})
at all points on the interval $[0,1]$. We are currently developing an improved
variational approach to address some of the shortcomings discussed above.

\section{Summary}
\label{sect:sum}

Cells live in a fluctuating environment in which signals and noise keep
bombarding the cell receptors~\cite{st02gomp,bar98bac,sm05inf}. Noisy signals
propagate inside the cell via microscopic chemical reaction events. 
Cells have evolved to adapt to or even exploit the seemingly
deleterious effect of fluctuations on signaling dynamics within a mesoscopic
size object. Thus, it is important to develop a qualitative picture, based on
mathematical modeling of stochastic chemical kinetics, of how signaling
networks process noisy signals. In this paper, we applied a variational principle
to the solution of the master equation which describes the noisy signal propagation. 

The essential difficulty associated with the master equation approach is the
enormous number of ODEs involved. To compactly encode information, we use a
QFT formulation in which the evolution of probability distributions is
governed by one ``quantum'' wave equation. We have explicitly demonstrated the
equivalence of the field theoretic formalism with the generating function
approach, greatly facilitating the practical application of the variational
technique proposed by Eyink~\cite{eyink96act}. We further examined the
significance of the variational principle in this context. According to our
previous investigation~\cite{lan1cas}, we suggest two novel classes of
time-dependent basis functions: one is in simple algebraic form and another is
in an integral convolution. These basis functions are key to the successful
application of the variational method to various signaling pathways. We
applied the new basis functions to describe stochastic signaling in 2-step,
3-step and 4-step enzymatic cascades and compared the obtained results with
alternative solution techniques.  The variational scheme presented here works
favorably in a large parameter range. It treats effectively both the small and
the large particle numbers, and is orders of magnitude faster to compute
compared with various Monte Carlo simulation algorithms.

However, the current scheme has also some limitations.  The resulting
evolution equations may be complicated and their derivation requires
considerable symbolic manipulation, somewhat ameliorated by using modern
computer algebra software. We also showed that the variational principle
itself in this context is not the most optimal. Despite these shortcomings,
the present variational approach may already be profitably applied to various
signal transduction pathways, allowing one to obtain quantitative and
semiquantitative solution to stochastic signaling dynamics in a broad range of
parameters. The technique may be further improved to extend its limits of
applicability, which is a work in progress.

\bibliography{../../biophys,../../nonlind}
\end{document}